# Incremental dynamics of prestressed viscoelastic solids and its applications in shear wave elastography

Yuxuan Jiang[1], Guo-Yang Li[2, *], Zhaoyi Zhang[1], Shiyu Ma[1], Yanping Cao[1, *],

Seok-Hyun Yun[3, 4]

[1] Institute of Biomechanics and Medical Engineering, AML, Department of Engineering Mechanics, Tsinghua University, Beijing 100084, PR China

[2] Department of Mechanics and Engineering Science, College of Engineering, Peking University, Beijing 100871, PR China

[3] Harvard Medical School and Wellman Center for Photomedicine, Massachusetts General Hospital, Boston, Massachusetts 02114, USA

[4] Harvard-MIT Division of Health Sciences and Technology, Cambridge, MA 02139, USA

\* Corresponding authors:
Guo-Yang Li (lgy@pku.edu.cn); Yanping Cao (caoyanping@tsinghua.edu.cn).

**Acknowledgements**

Yanping Cao acknowledges support from the National Natural Science Foundation of China (Grants Nos. 11972206 and 11921002). Guo-Yang Li acknowledges the financial support from the National Natural Science Foundation of China (Grant No. 12472176) and the Fundamental Research Funds for the Central Universities, Peking University. Seok-Hyun Yun acknowledges funding from the U.S. National Institutes of Health via grants R01EY033356, and R01EY034857.

**Abstract:**

Shear wave elastography (SWE) is a promising imaging modality for mechanical characterization of tissues, offering biomarkers with potential for early and precise diagnosis. While various methods have been developed to extract mechanical parameters from shear wave characteristics, their relationships in viscoelastic materials under prestress remain poorly understood. Here, we present a generalized incremental dynamics theory for finite-strain viscoelastic solids. The theory derives small-amplitude viscoelastic wave motions in a material under static pre-stress. The formalism is compatible with a range of existing constitutive models, including both hyperelasticity and viscoelasticity—such as the combination of Gasser-Ogden-Holzapfel (GOH) and Kelvin-Voigt fractional derivative (KVFD) models used in this study. We validate the theory through experiments and numerical simulations on prestressed soft materials and biological tissues, using both optical coherence elastography and ultrasound elastography. The theoretical predictions closely match experimental dispersion curves over a broad frequency range and accurately capture the effect of prestress. Furthermore, the framework reveals the relationships among shear wave phase velocity, attenuation, and principal stresses, enabling prestress quantification in viscoelastic solids without prior knowledge of constitutive parameters. This generalized acousto-viscoelastic formalism is particularly well-suited for high-frequency, high-resolution SWE in tissues under prestress.

***Keywords:*** Acoustoelasticity; Viscoelasticity; Soft tissue rheology; Elastic waves; Elastography.

# 1 Introduction

The emergence of shear wave elastography (SWE) technologies has made the mechanical properties of soft biological tissues available as a biomarker that holds the promise to address unmet needs in early and precise diagnosis of diseases, such as staging liver fibrosis (Ferraioli et al., 2015) and assessing breast tumor (Barr et al., 2015). In SWE, traveling elastic waves over a limited frequency band are generated by means of noninvasive stimuli and then visualized using medical imaging modalities, such as ultrasound (Gennisson et al., 2013), magnetic resonance imaging (Mariappan et al., 2010), and optical coherence tomography (Kennedy et al., 2013). The speeds of the traveling elastic waves can offer a clear contrast for normal and diseased tissues as they are primarily determined by mechanical properties of the tissues that can be altered by pathology. To quantitatively infer the mechanical properties of soft tissues with traveling waves, wave theories relying on constitutive models that are able to describe the deformation behaviors of soft tissues in vivo are necessary (Cao et al., 2019; Li & Cao, 2017).

It is well recognized that most soft biological tissues are viscoelastic and subject to prestress (Chen et al., 2010; Mammoto & Ingber, 2010). The viscoelastic deformation behaviors and the presence of prestress play essential roles in their normal physiological functions and may be altered by diseases (Cyron & Humphrey, 2017; Sack et al., 2013). To infer the viscoelastic properties of soft tissues *in vivo* with SWE, different viscoelastic models have been used to characterize the features of wave dispersion (Sack et al., 2013; Zhang et al., 2021; Zhou & Zhang, 2018). Notably, existing data suggest that the viscoelasticity-caused wave dispersions in soft tissues can be well predicted by the Kelvin-Voigt fractional derivative model (KVFD, or the power-law rheological model) (Bonfanti et al., 2020), upon which an SWE method to probe viscoelastic properties of soft tissues in a broad frequency range can be developed (Parker et al., 2019; Poul et al., 2022). Besides the measurement of viscoelastic properties *in vivo*, SWE is also promising in probing the prestress in soft tissues. Prestresses exist in load-bearing tissues such as arteries and corneas, and can dramatically alter shear wave speeds due to the nonlinear stiffening behavior of soft biological tissues (Couade et al., 2010; Li et al., 2022b). The small-amplitude shear wave utilized in SWE can be modelled as incremental motions superposed on the large deformation introduced by the prestress, coined as incremental dynamic theory (Destrade, 2015; Ogden, 2007), which forms the theoretical basis to develop an SWE method to infer prestress (Zhang et al., 2023).

It is of notice that the effects of viscoelasticity and prestress on wave motions in soft biological tissues have been investigated separately and corresponding SWE methods to infer either viscoelastic properties (Parker et al., 2019; Poul et al., 2022; Zheng et al., 2021) or prestresses (Zhang et al., 2023) have been developed in parallel. However, imaging the viscoelasticity of soft tissues with shear waves can suffer from the effect of prestress on wave dispersion; meanwhile, neglecting the viscoelasticity in inferring prestress from shear wave speeds may result in significant errors. To address these fundamental issues, an incremental dynamics theory for prestressed viscoelastic solids within the framework of continuum mechanics has been suggested in this study. Different from the theories relying on the Kelvin–Voigt model (Colonnelli et al., 2013; Destrade et al., 2009; Saccomandi, 2005) or that taking the Prony series as the relaxation kernel function (Berjamin & De Pascalis, 2022; Parnell & De Pascalis, 2019), the KVFD model is adopted to describe the power-law rheology of soft tissues observed in experiments. Based on such a theoretical framework, several analytical solutions to predict shear waves, surface waves and guided waves in prestressed viscoelastic solids are presented. Particular attention is paid to the effect of viscoelasticity and prestress on wave dispersion and attenuation, which not only helps quantify the influence of material viscoelasticity and prestress on elastic wave propagation in soft tissues, but also leads to a simple SWE method to infer prestress in a viscoelastic solid.

This paper is organized as follows. In section 2, we derive the incremental equation of motions for prestressed viscoelastic materials following the framework of elastic incremental dynamics (Destrade, 2015; Ogden, 2007). In particular, we introduce the fractional order derivative of the stress (which is equivalent to the KVFD model) to capture the power law rheology of biological tissues. In section 3, the analytical solutions of dispersion and attenuation of plane shear waves, surface/interface waves and Lamb waves in uniformly prestressed viscoelastic solids are derived. By applying these theoretical solutions, we discuss the effects of prestress and material viscoelasticity on phase velocity and dissipation. In section 4, to verify theoretical dispersion relations and demonstrate their usefulness in practical measurements, we perform SWE on soft artificial materials and *ex vivo* soft biological tissues, and analyze the experiments with the proposed theory. In section 5, we present an identity that relates biaxial prestress to the biaxial wave velocities of plane shear waves, enabling direct stress measurement through elastic wave motions. This finding expands upon the work of acoustoelastic imaging method to probe stress in elastic soft materials (Zhang et al., 2023). We verify the proposed method by finite element analysis. Finally, in section 6 we give the concluding remarks.

## 2   Incremental dynamics of prestressed viscoelastic solids

### 2.1 Overview of incremental dynamics

Here we briefly revisit the theoretical background of incremental dynamics that have been developed for pure elastic materials. Readers are referred to Ogden (2007) and Destrade (2015) for more details. The definition of notations is consistent with those used in the reference work (Destrade, 2015; Ogden, 2007) (see details in Supplementary Note 1).

*2.1.1   Kinematics*

Fig. 1a and 1b illustrate the kinematics of elastic and viscoelastic materials, respectively. We denote the initial, deformed and incremental configurations with $\mathcal{B}_0$, $\mathcal{B}$ and $\mathcal{B}'$, respectively. The coordinates corresponding to $\mathcal{B}_0$, $\mathcal{B}$ and $\mathcal{B}'$ are denoted as $X$, $x$ and $x'$, respectively. While the deformation process is the same for both elastic and viscoelastic materials, their stress states and material moduli differ. For the viscoelastic materials, we assume the stress relaxation involved in the deformation from $\mathcal{B}_0$ to $\mathcal{B}$ has been fully developed, so $\mathcal{B}$ is in equilibrium. $\mathcal{B}'$ is an infinitesimal perturbation from $\mathcal{B}$. The incremental motion is denoted by $u$, i.e., $x' = x + u$. The incremental stress in the material depends on the frequency of the incremental motion. The deformation gradient tensors for $\mathcal{B}_0 \to \mathcal{B}$ and $\mathcal{B}_0 \to \mathcal{B}'$ are denoted by $F$ and $F'$, respectively. Using the chain rule, we have

$$F' = \frac{\partial x'}{\partial x}\frac{\partial x}{\partial X} = (I + \Gamma)F \equiv F + \hat{F} ,\qquad(1)$$

where $\Gamma \equiv \partial u / \partial x$. With the incompressible constraint for soft materials, we have $J = \det(F) = 1$ and $J' = \det(F') = 1$.

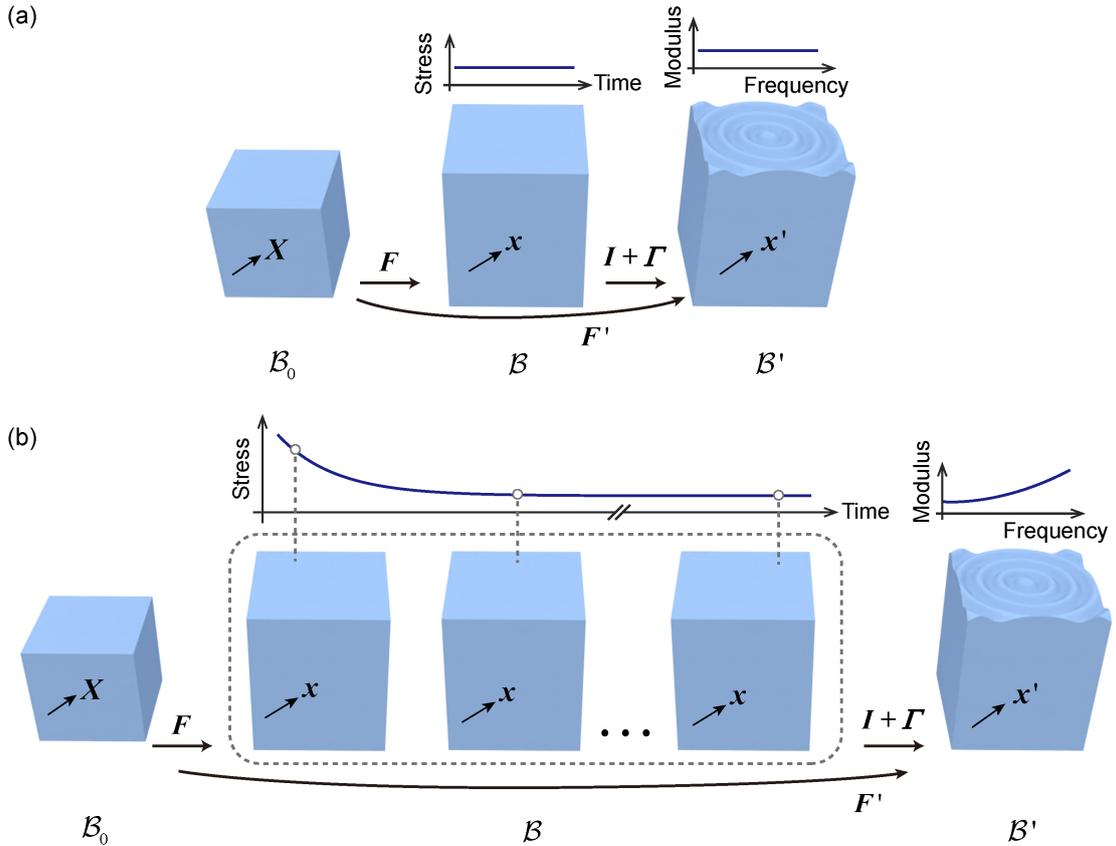

**Figure 1.** Configurations involved in the theoretical analysis, including initial ($\mathcal{B}_0$), deformed ($\mathcal{B}$) and incremental states ($\mathcal{B}'$). (a) Kinematics of elastic materials. (b) Kinematics of viscoelastic materials. For the viscoelastic materials, the stress is fully

relaxed at $\mathcal{B}$; the complex modulus increases with the increase of wave frequency at $\mathcal{B}'$.

*2.1.2 Incremental dynamics*

The equation of motion on $\mathcal{B}$ is $\text{Div}\boldsymbol{S} = \rho\ddot{\boldsymbol{x}}$, where $\boldsymbol{S}$ is the nominal stress. Nominal stress $\boldsymbol{S}$ is related to the Cauchy stress $\boldsymbol{\sigma}$ and PK-II stress $\boldsymbol{T}$ by $\boldsymbol{\sigma} = \boldsymbol{FS}$ and $\boldsymbol{T} = \boldsymbol{SF}^{-\text{T}}$, respectively. The notation of divergence 'Div' is defined with respect to $\boldsymbol{X}$. Similarly, the equation of motion on $\mathcal{B}'$ is $\text{Div}\boldsymbol{S}' = \rho\ddot{\boldsymbol{x}}'$. By taking the difference of the two equations, we get the equation that governs the incremental motions

$$\text{Div}\hat{\boldsymbol{S}} = \rho\ddot{\boldsymbol{u}}, \tag{2}$$

where $\hat{\boldsymbol{S}} = \boldsymbol{S}' - \boldsymbol{S}$. The notation '^' indicates the increment of a quantity. Taking the incremental form of $\boldsymbol{S} = \boldsymbol{F}^{-1}\boldsymbol{\sigma}$, we get $\hat{\boldsymbol{S}} = \boldsymbol{F}^{-1}\left(\hat{\boldsymbol{\sigma}} - \hat{\boldsymbol{F}}\boldsymbol{F}^{-1}\boldsymbol{\sigma}\right)$. By pushing forward of Eq. (2), we obtain the incremental equation of motion

$$\text{div}\boldsymbol{\Sigma} = \rho\ddot{\boldsymbol{u}}, \tag{3}$$

where

$$\boldsymbol{\Sigma} \equiv \boldsymbol{F}\hat{\boldsymbol{S}} = \hat{\boldsymbol{\sigma}} - \boldsymbol{\Gamma}\boldsymbol{\sigma}. \tag{4}$$

The divergence 'div' is computed with respect to $\boldsymbol{x}$. By taking the incremental form of $\boldsymbol{\sigma} = \boldsymbol{F}\boldsymbol{T}\boldsymbol{F}^{\text{T}}$, we obtain the relationship between incremental Cauchy stress $\hat{\boldsymbol{\sigma}}$ and incremental PK-II stress $\hat{\boldsymbol{T}}$ as

$$\hat{\boldsymbol{\sigma}} = \boldsymbol{\Gamma}\boldsymbol{\sigma} + \boldsymbol{\sigma}\boldsymbol{\Gamma}^{\text{T}} + \boldsymbol{F}\hat{\boldsymbol{T}}\boldsymbol{F}^{\text{T}}. \tag{5}$$

Inserting Eq. (5) into (4), we get

$$\boldsymbol{\Sigma} = \boldsymbol{\sigma}\boldsymbol{\Gamma}^{\text{T}} + \boldsymbol{F}\hat{\boldsymbol{T}}\boldsymbol{F}^{\text{T}}. \tag{6}$$

**2.2 Constitutive models of viscoelastic materials**

Nonlinear viscoelastic theories represent an evolving field, and various viscoelastic models have been developed in the literature (Wineman, 2009). In this work, two types of nonlinear viscoelastic models are employed to develop the acoustoelastic theory: the quasi-linear viscoelastic (QLV) model (De Pascalis et al., 2014; Fung, 1993) with the Prony series as the relaxation kernel function, and the Kelvin-Voigt fractional derivative (KVFD) model (Adolfsson & Enelund, 2003; Nordsletten et al., 2021).

Following the assumption introduced by Simo (1987), the hydrostatic and deviatoric parts of the second Piola-Kirchhoff (PK-II) stress in the incompressible viscoelastic solids exhibit independent viscoelastic responses. The PK-II stress $T$ is decomposed as

$$T = -q\boldsymbol{C}^{-1} + \boldsymbol{T}_D, \tag{7}$$

where $q$ denotes an incompressible term. $\boldsymbol{C} = \boldsymbol{F}^{\mathrm{T}}\boldsymbol{F}$ denotes the right Cauchy-Green tensor. $\boldsymbol{T}_D$ ( $= \mathrm{Dev}(\boldsymbol{T})$ ) is the deviatoric stress, where $\mathrm{Dev}(\cdot) = (\cdot) - ([\cdot]:\boldsymbol{C})\boldsymbol{C}^{-1}/3$ denotes the deviatoric operator in the Lagrangian description.

### 2.2.1 Quasi-linear viscoelastic model with Prony series

The QLV model assumes that the current stress depends solely on the history of stress (Fung, 1993). The key limitations of this model are its inability to capture the initial stress-dependent behavior during stress relaxation, and its insufficiency in characterizing creep behavior. Nevertheless, it remains a useful approximation for characterizing the viscoelastic behavior of biological soft tissues in many cases (De Pascalis et al., 2018). According to the QLV model assumption, the deviatoric stress can be expressed by a hereditary integral

$$\boldsymbol{T}_D = \mathcal{G} * \dot{\boldsymbol{T}}_D^e = \int_0^t \mathcal{G}(t-s) \cdot \frac{\partial \boldsymbol{T}_D^e(s)}{\partial s} \mathrm{d}s, \tag{8}$$

where the notation '$*$' denotes the convolution operator. $t$ denotes time. The material is assumed to be stress-free for negative times. The deviatoric part of elastic stress $\boldsymbol{T}_D^e$ is defined as $\boldsymbol{T}_D^e = \mathrm{Dev}(\boldsymbol{T}^e)$, where $\boldsymbol{T}^e$ is derived from the strain energy function $W$ by $\boldsymbol{T}^e = (\partial W / \partial \boldsymbol{F})\boldsymbol{F}^{-\mathrm{T}}$. $\mathcal{G}(t)$ is the relaxation kernel function. The relaxation function of the Prony series model is defined as

$$\mathcal{G}(t) = 1 - \sum_{k=1}^{n} g_k \left[1 - \exp(-t/\tau_k)\right], \tag{9}$$

where $g_k$ and $\tau_k$ ($k = 1, 2, \ldots, n$) denote the $k$-th order relaxation magnitude and characteristic time, respectively. Inserting Eq. (9) into Eq. (8), we get (Berjamin & De Pascalis, 2022)

$$\boldsymbol{T}_D = \boldsymbol{T}_D^e - \sum_{k=1}^{n} \boldsymbol{T}_k^v, \tag{10}$$

where $\boldsymbol{T}_k^v \equiv (g_k / \tau_k) \int_0^t e^{-(t-s)/\tau_k} \boldsymbol{T}_D^e(s) \mathrm{d}s$. $\boldsymbol{T}_k^v$ can be regarded as internal variables and their evolution equation is

$$\tau_k \dot{\boldsymbol{T}}_k^v = g_k \boldsymbol{T}_D^e - \boldsymbol{T}_k^v. \tag{11}$$

*2.2.2   Kelvin-Voigt fractional derivative model*

The KVFD model employed in this work assumes that the current viscous stress depends on the history of stress, which shares a common fundamental assumption with the QLV model. The viscous stress $\boldsymbol{T}_D$ is determined by (Capilnasiu et al., 2020; Nordsletten et al., 2021)

$$\boldsymbol{T}_D = \boldsymbol{T}_D^e + \boldsymbol{T}^v, \tag{12}$$

where $\boldsymbol{T}^v$ is the fractional derivation of $\boldsymbol{T}_D^e$, defined as

$$\boldsymbol{T}^v \equiv \eta \frac{\mathrm{d}^{\beta_0} \boldsymbol{T}_D^e}{\mathrm{d}t^{\beta_0}}, \tag{13}$$

where $\beta_0$ denotes the fractional order ($0 < \beta_0 < 1$; when $\beta_0 = 0$ the model reduces to a purely elastic material; when $\beta_0 = 1$, it corresponds to the classical Kelvin-Voigt viscoelastic model). $\eta$ reflects the relative contribution of viscosity to elasticity in the material (dimension $[\mathrm{s}^{\beta_0}]$).

While the two models aforementioned will be discussed in this study, we are primarily interested in the KVFD model, which has been proved to match the experimental data for soft biological tissues better (Bonfanti et al., 2020; Poul et al., 2022). It should be noted that the elastic stress $\boldsymbol{T}_D^e$ in Eq. (12) represents the long-term (fully relaxed) elastic response. In contrast, the term $T_D^e$ in Eq. (10) corresponds to the instantaneous elastic response.

## 2.3 Incremental motions of viscoelastic solids

Based on the framework of incremental dynamics introduced in Sec 2.1, we aim to derive the incremental motion equation of viscoelastic solids. Since the hydrostatic and deviatoric stresses are introduced in the viscoelastic soft biological materials, we reformulate the form of incremental stress $\Sigma$ as follows. Taking the incremental form of Eq. (7), we have

$$F\hat{T}F^{\mathrm{T}} = -\hat{q}I + q\Gamma + q\Gamma^{\mathrm{T}} + F\hat{T}_D F^{\mathrm{T}}, \qquad (14)$$

where $\hat{q}$ and $\hat{T}_D$ are increments of $q$ and $T_D$, respectively. Inserting Eqs. (14) and $\sigma = -qI + \sigma_D$ into Eq. (6), we get

$$\Sigma = -\hat{q}I + q\Gamma + \sigma_D \Gamma^{\mathrm{T}} + F\hat{T}_D F^{\mathrm{T}}. \qquad (15)$$

In Eq. (15), $\sigma_D$ denotes the fully relaxed deviatoric Cauchy stress at $\mathcal{B}$. $\hat{T}_D$ denotes the incremental deviatoric PK-II stress at $\mathcal{B}'$. For Prony series model, $\sigma_D$ is

$$\sigma_D = \left(1 - \sum_{k=1}^{n} g_k\right)\sigma_D^e = (1-g)\sigma_D^e, \qquad (16)$$

where $g \equiv \sum_{k=1}^{n} g_k$. We further assume the incremental motion is harmonic (angular frequency $\omega = 2\pi f$); therefore, all the incremental quantities admit a harmonic formulation. Then according to Eq. (11) we get $\hat{T}_k^v = \dfrac{g_k}{1+i\omega\tau_k}\hat{T}_D^e$, which, together with Eq. (10), helps to obtain

$$\hat{T}_D = \left(1 - \sum_{k=1}^{n} \frac{g_k}{1+i\omega\tau_k}\right)\hat{T}_D^e. \qquad (17)$$

Inserting Eqs. (16) and (17) into Eq. (15), we get

$$\Sigma = -\hat{q}I + q\Gamma + (G-\Omega)\sigma_D^e \Gamma^{\mathrm{T}} + GF\hat{T}_D^e F^{\mathrm{T}}, \qquad (18)$$

where $G$ and $\Omega$ are defined as

$$G \equiv 1 - \sum_{k=1}^{n} \frac{g_k}{1 + i\omega\tau_k}, \tag{19}$$

and

$$\Omega \equiv G - (1 - g) = \sum_{k=1}^{n} g_k \frac{i\omega\tau_k}{1 + i\omega\tau_k}, \tag{20}$$

respectively.

For the KVFD model, according to Eqs. (12) and (13), we obtain $\sigma_D = \sigma_D^e$ as the relaxed stress and $\hat{T}_D^v = \eta(i\omega)^{\beta_0} \hat{T}_D^e$ as the harmonic incremental stress. As a result, the incremental stress $\Sigma$ for the KVFD model can be expressed in the same form as Eq. (18), with only the parameters $G$ and $\Omega$ replaced by

$$G \equiv 1 + \eta(i\omega)^{\beta_0}, \tag{21}$$

and,

$$\Omega \equiv G - 1 = \eta(i\omega)^{\beta_0}. \tag{22}$$

We proceed to introduce the constitutive model into Eq. (18) to eliminate elastic stresses $\sigma_D^e$ and $\hat{T}_D^e$. Deviatoric elastic stresses $\sigma_D^e$ and $T_D^e$ are defined by

$$\sigma_D^e \equiv \sigma^e - \frac{1}{3}(\sigma^e : I)I \equiv F\frac{\partial W}{\partial F} - QI, \tag{23}$$

and

$$T_D^e \equiv T^e - \frac{1}{3}(T^e : C)C^{-1} \equiv \frac{\partial W}{\partial F}F^{-T} - QC^{-1}, \tag{24}$$

where $Q \equiv (T^e : C)/3 = (\sigma^e : I)/3$. Taking the incremental form of Eq. (24), we get

$$F\hat{T}_D^e F^T = F\frac{\partial^2 W}{\partial F \partial F}\Gamma F - F\frac{\partial W}{\partial F}\Gamma^T - \hat{Q}I + Q(\Gamma + \Gamma^T), \tag{25}$$

where $\hat{Q}$ denotes the increment of $Q$. Inserting Eqs. (23) and (25) into Eq. (18), we get

$$\Sigma = -\hat{q}I + q\Gamma - G\hat{Q}I + GQ\Gamma + G\mathcal{A}_0\Gamma - \Omega\sigma_D^e\Gamma^T, \tag{26a}$$

and its component form is

$$\Sigma_{ji} = -\hat{q}\delta_{ji} + qu_{j,i} - G\hat{Q}\delta_{ji} + GQu_{j,i} + G\mathcal{A}_{0jikl}u_{l,k} - \Omega\sigma_{Djk}^e u_{i,k}, \tag{26b}$$

where $i, j, k, l \in \{1, 2, 3\}$, denoting components along the three Cartesian coordinate directions— $x_1$, $x_2$, and $x_3$. The subscript with a comma denotes partial differentiation with respect to the corresponding variable. $\mathcal{A}_0$ is the fourth-order Eulerian elasticity tensor with components $\mathcal{A}_{0jikl} = \dfrac{\partial^2 W}{\partial F_{iM} \partial F_{lN}} F_{jM} F_{kN}$ ($M, N \in \{1, 2, 3\}$). It should be noted that the stresses (e.g., $Q$ and $\sigma_D^e$) and the elasticity tensor (also referred to as the incremental stiffness) $\mathcal{A}_0$ in Eq. (26) refer to the instantaneous values for the QLV model, and the long-term values for the KVFD model.

Finally, inserting Eq. (26b) into Eq. (3), we get the incremental motion equation for the uniformly prestressed viscoelastic solids:

$$G\mathcal{A}_{0jikl} u_{l,jk} - \hat{q}_{,i} - G\hat{Q}_{,i} - \Omega \sigma^e_{Djk} u_{i,jk} = \rho u_{i,tt}, \qquad (27)$$

coupled with the incompressible constraint

$$u_{i,i} = 0. \qquad (28)$$

To get Eq. (27), we have used homogeneous deformation conditions $q_{,i} = 0$, $Q_{,i} = 0$, and $\mathcal{A}_{0jikl,j} = 0$, and the incompressible constraint $\Gamma_{ji,j} = 0$.

For pure elastic solids, we have $G = 1$ and $\Omega = 0$. The PK-II stress recovers to $\boldsymbol{T} = -q\boldsymbol{C}^{-1} + \boldsymbol{T}_D^e = -p\boldsymbol{C}^{-1} + \boldsymbol{T}^e$, where $p \equiv q + Q$ denotes the Lagrange multiplier. Taking the increments of the quantities yields: $\hat{p} = \hat{q} + \hat{Q}$. Then Eq. (27) reduces to the equation of incremental motions for elastic solids (Ogden, 2007)

$$\mathcal{A}_{0jikl} u_{l,jk} - \hat{p}_{,i} = \rho u_{i,tt}. \qquad (29)$$

## 3  Small-amplitude waves in prestressed viscoelastic solids

In this section, the incremental dynamic theory is implemented to study the small-amplitude wave motions in uniformly prestressed viscoelastic solids. Three types of elastic waves frequently involved in SWE are discussed, i.e., bulk shear wave, surface/interface wave, and Lamb wave. Analytical dispersion equations for the waves are derived and the verification of the results by finite element analysis is provided in supplementary materials (see details in Supplementary Note 2).

**3.1 Plane Shear wave**

We consider plane shear waves propagating in the $x_1 - x_2$ plane with in-plane polarization; therefore, only the displacement components $u_1$ and $u_2$ are nonzero. Taking $u_1$ and $u_2$ into the wave motion equation (27), we get

$$G\left(\mathcal{A}_{01111}u_{1,11} + \mathcal{A}_{01122}u_{2,12} + \mathcal{A}_{02121}u_{1,22} + \mathcal{A}_{02112}u_{2,21}\right) \\ -\Omega\left(\sigma^e_{D11}u_{1,11} + \sigma^e_{D22}u_{1,22}\right) - \hat{q}_{,1} - G\hat{Q}_{,1} = \rho u_{1,tt} \quad (30)$$

and

$$G\left(\mathcal{A}_{01212}u_{2,11} + \mathcal{A}_{02222}u_{2,22} + \mathcal{A}_{01122}u_{1,12} + \mathcal{A}_{01221}u_{1,12}\right) \\ -\Omega\left(\sigma^e_{D11}u_{2,11} + \sigma^e_{D22}u_{2,22}\right) - \hat{q}_{,2} - G\hat{Q}_{,2} = \rho u_{2,tt} \quad (31)$$

Eliminating $\hat{q}$ and $\hat{Q}$ in Eqs. (30) and (31), and introducing a stream function $\psi(x_1, x_2, t)$ — which satisfies $u_1 = \psi_{,2}$ and $u_2 = -\psi_{,1}$ — to take place of $u_1$ and $u_2$, we get

$$G\left[\alpha\psi_{,1111} + \gamma\psi_{,2222} + 2\beta\psi_{,1122}\right] \\ -\Omega\left[\sigma^e_{D11}\psi_{,1111} + \sigma^e_{D22}\psi_{,2222} + \left(\sigma^e_{D11} + \sigma^e_{D22}\right)\psi_{,1122}\right] = \rho\left(\psi_{,22tt} + \psi_{,11tt}\right) \quad (32)$$

where $\alpha \equiv \mathcal{A}_{01212}$, $\gamma \equiv \mathcal{A}_{02121}$, $\beta \equiv (\mathcal{A}_{01111} + \mathcal{A}_{02222} - 2\mathcal{A}_{01122} - 2\mathcal{A}_{01221})/2$. In the derivation of Eq. (32), we have used the major symmetry of the tensor $\mathcal{A}_0$, i.e. $\mathcal{A}_{0jikl} = \mathcal{A}_{0klji}$. As a supplementary discussion, we provide a relationship between stresses and incremental parameters. This relationship is useful for calculating dispersion relations in practice and will also be used in Sec. 5. With the help of the identity $\sigma^e_{ii} = \mathcal{A}_{0ijij} - \mathcal{A}_{0jiij}$ ($i \neq j$, no summation) (Destrade, 2015), where $\sigma^e$ has been defined in Eq. (23), the volumetric stress $Q$ is related to $\mathcal{A}_{0jikl}$ by

$$Q = \frac{1}{3}\left(\mathcal{A}_{01212} + \mathcal{A}_{02121} + \mathcal{A}_{03232} - 2\mathcal{A}_{01221} - \mathcal{A}_{02332}\right). \quad (33)$$

The deviatoric stresses $\sigma^e_{D11}$ and $\sigma^e_{D22}$ are related to $\mathcal{A}_{0jikl}$ by

$$\sigma^e_{D11} = \frac{2}{3}\mathcal{A}_{01212} - \frac{1}{3}\mathcal{A}_{01221} - \frac{1}{3}\mathcal{A}_{02121} - \frac{1}{3}\mathcal{A}_{03232} + \frac{1}{3}\mathcal{A}_{02332}, \quad (34)$$

and

$$\sigma^e_{D22} = \frac{2}{3}\mathcal{A}_{02121} - \frac{1}{3}\mathcal{A}_{01221} - \frac{1}{3}\mathcal{A}_{01212} - \frac{1}{3}\mathcal{A}_{03232} + \frac{1}{3}\mathcal{A}_{02332}, \quad (35)$$

respectively.

For plane shear waves propagating at an angle $\theta$ with respect to the $x_1$ axis, the stream function can be expressed as $\psi = \psi_0 \exp[ik(x_1 \cos\theta + x_2 \sin\theta)]\exp(-i\omega t)$, where $\psi_0$ denotes the amplitude, $k$ is the wavenumber. Inserting $\psi$ into Eq. (32), we obtain the general solution of plane shear waves in a prestressed viscoelastic material:

$$\begin{aligned}\rho\mathcal{C}^2 &= (G\alpha - \Omega\sigma^e_{D11})\cos^4\theta \\ &+ (G\gamma - \Omega\sigma^e_{D22})\sin^4\theta + [2G\beta - \Omega(\sigma^e_{D11} + \sigma^e_{D22})]\sin^2\theta\cos^2\theta\end{aligned}, \quad (36)$$

where $\mathcal{C}$ ($=\omega/k$) denotes the complex wave velocity. The phase velocity $c$ and wave dissipation factor $d$ then can be calculated by

$$c \equiv \frac{\omega}{\text{Re}(k)} = [\text{Re}(\mathcal{C}^{-1})]^{-1}, \quad (37)$$

and

$$d \equiv \frac{\text{Im}(k^2)}{\text{Re}(k^2)} = \frac{\text{Im}(\mathcal{C}^{-2})}{\text{Re}(\mathcal{C}^{-2})}. \quad (38)$$

Firstly, for a comparison with the literature result (Berjamin & De Pascalis, 2022), we consider a plane shear wave propagating along the $x_1$ axis (i.e. $\theta = 0$). In the referenced work, the Mooney-Rivlin model was adopted, with the strain energy function given by $W = C_{10}(I_1 - 3) + C_{01}(I_2 - 3)$, where $C_{10}$ and $C_{01}$ are two constitutive parameters, $I_1$ and $I_2$ are two invariants (see details in Supplementary Note 7.2). The material viscoelasticity is described by the one-term Prony series (including viscoelastic parameters $g_1$ and $\tau_1$). The material is subjected to a uniaxial stretch along the $x_2$ direction; therefore, the deformation gradient tensor is $F = \text{diag}(\lambda^{-1/2}, \lambda, \lambda^{-1/2})$. Substituting the above relations into Eq. (36), the plane shear wave is obtained as

$$\rho \mathcal{C}^2 = \left(1 - \frac{g_1}{1+i\omega\tau_1}\right)\left(2C_{10}\lambda^{-1} + 2C_{01}\lambda^{-2}\right) \\ - \frac{ig_1\omega\tau_1}{1+i\omega\tau_1}\left[\frac{2}{3}C_{10}\left(\lambda^{-1} - \lambda^2\right) + \frac{2}{3}C_{01}\left(\lambda^{-2} - \lambda\right)\right]. \tag{39}$$

It can be verified that Eq. (39) is consistent with the solution in the referenced work (see details in Supplementary Note 3).

In the following, we discuss the effect of prestress on plane shear waves. For simplification, we consider a neo-Hookean material with the strain energy function $W = \mu(I_1 - 3)/2$, where $\mu$ denotes the small-strain and long-term shear modulus. An in-plane uniaxial stretch is applied along the $x_1$ direction, with the deformation gradient tensor $F = \text{diag}(\lambda, \lambda^{-1}, 1)$. By substituting the above relations into Eq. (36), the expressions for plane shear waves in a viscoelastic neo-Hookean material can be obtained (see Eq. (S18) in Supplementary Note 4).

For the KVFD model, the complex wave velocities of plane shear waves propagating along the $x_1$-axis and $x_2$-axis are (see details in Supplementary Note 4)

$$\rho \mathcal{C}_1^2 = \mu\lambda^2 + \frac{1}{3}\mu\eta(i\omega)^{\beta_0}\left(\lambda^2 + \lambda^{-2} + 1\right), \tag{40a}$$

and

$$\rho \mathcal{C}_2^2 = \mu\lambda^{-2} + \frac{1}{3}\mu\eta(i\omega)^{\beta_0}\left(\lambda^2 + \lambda^{-2} + 1\right), \tag{40b}$$

respectively. We define the dimensionless phase velocity and frequency as $\tilde{c} \equiv c/\sqrt{\mu/\rho}$ and $\tilde{f} \equiv f\eta^{1/\beta_0}$, respectively. By applying Eqs. (40a) and (40b), Fig. 2a clearly shows that the phase velocities increase with the frequency. In the high frequency regime, the dispersion relations follow a power law $\tilde{c} \propto \tilde{f}^{\beta_0/2}$. Our first insight is that the effect of prestress on phase velocity will be quenched in the high-frequency regime. Taking the $\lambda = 1.67$ as an example, in the low frequency regime ($\tilde{f} \to 0$) the stretch increases the phase velocity to $1.67\sqrt{\mu/\rho}$ for $\theta = 0$, whereas decreases the phase velocity to $0.6\sqrt{\mu/\rho}$ for $\theta = \pi/2$. As $\tilde{f} \to +\infty$, the difference between the two phase velocities gradually diminish. Figure 2b presents the angular distribution of normalized phase velocity at different frequencies ($\tilde{f} = 10^{-4}$, $10^{-1}$, $10^{3}$). At low frequency regime ($\tilde{f} = 10^{-4}$), the phase velocity is direction dependent, with a shape that will result in an elliptical group velocity curve (Zhang et al., 2023). As the frequency increases, the phase velocity gradually becomes isotropic. The quenching of the acoustoelastic effect in the high-frequency regime is primarily attributed to the KVFD model, in which the mechanical response of the spring—modifiable by prestress—becomes negligible at high frequencies. In contrast to the KVFD model, the Prony series model discussed later can retain the acoustoelastic effect due to the presence of a spring that is not arranged in parallel with a dashpot (i.e., the standard linear solid).

Figure 2c shows the variation of the dissipation factor ($d$) with respect to stretch ratio ($\lambda$). Remarkably, the dissipation factor decreases with stretch applied along the wave propagation direction across a broad frequency range, while in the transverse direction—perpendicular to the stretch— $d$ increases due to the compressive deformation introduced by Poisson's effect. The tunability of the dissipation factor by stress could serve as a useful strategy, particularly in the design of soft elastic waveguides with low dissipation, as soft materials typically exhibit significant energy loss in the high-frequency regime.

For comparison, Fig. 3 shows the results obtained using the QLV Prony series model (The expressions for plane shear waves are provided in Supplementary Note 4). Here, we define the dimensionless phase velocity and frequency as $\tilde{c} \equiv c/\sqrt{\mu/\rho}$ and $\tilde{f} \equiv f\tau$, respectively. Differently, the dispersion relations reach plateaus that still show dependence on $\lambda$ in the high-frequency regime ($\tilde{f} > 1$), in line with the anisotropic phase velocity profile shown in Fig. 3b ($\tilde{f} = 10^3$). The dispersion plateaus in the high-frequency regime are unlikely to occur in soft biological tissue, making the Prony series model less suitable than the KVFD model, especially when broad-band frequency data are involved (Feng et al., 2023; Hang et al., 2022; Parker et al., 2019). Therefore, in the remaining part of this paper, we will focus on the KVFD model. Figure 3c suggests that the dissipation factor decreases with increasing stretch. Different from the KVFD model, the dissipation factor reaches a maximum near $\tilde{f} = 1/2\pi$ and decreases to zero when $\tilde{f} \to +\infty$.

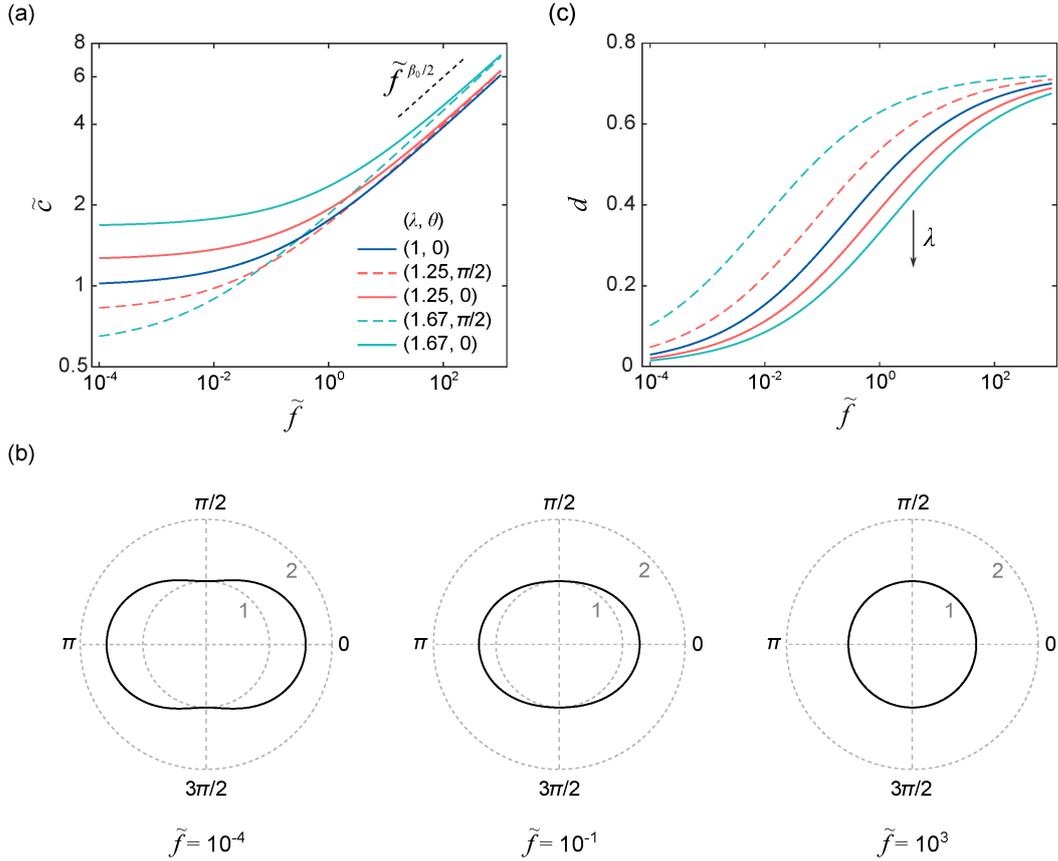

**Figure 2.** Effect of prestress on plane shear waves in the KVFD material. (a) Dimensionless dispersion relation. $\lambda$, stretch ratio, $\theta$, angle of the wave propagation direction. (b) Angular distribution of normalized phase velocity. $\lambda = 1.25$. (c) Dimensionless dissipation factor.

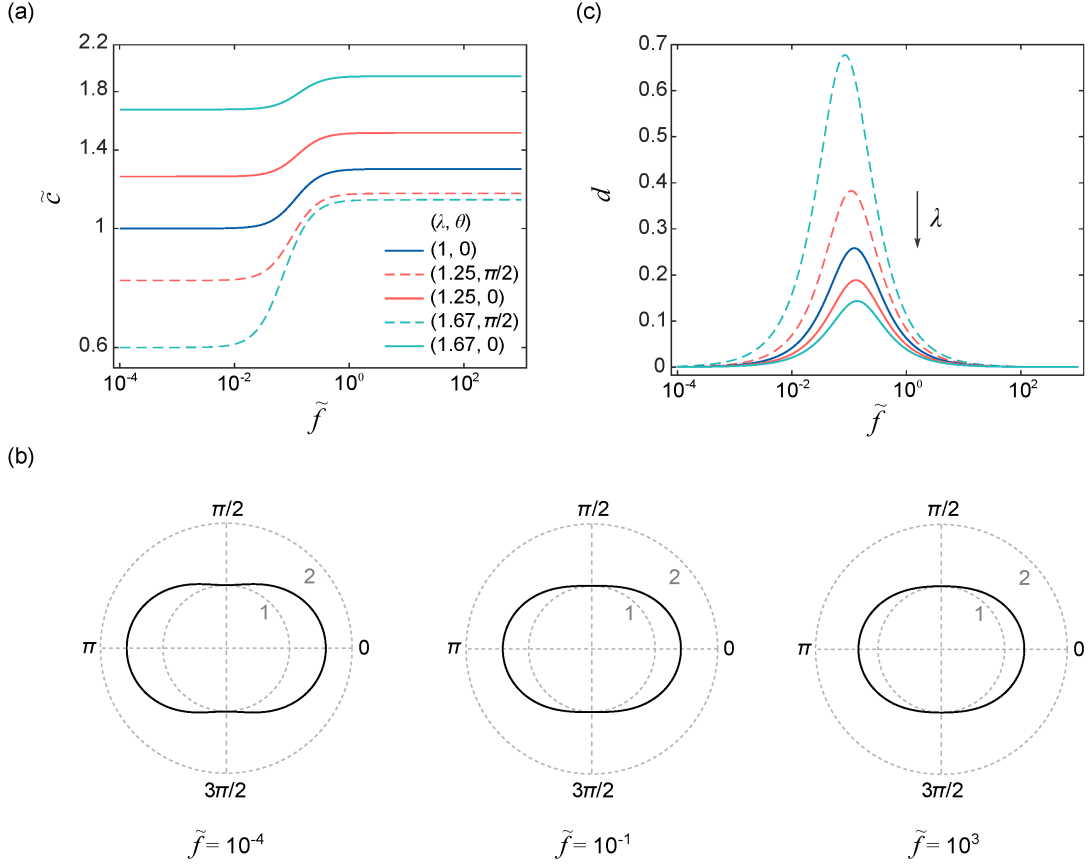

**Figure 3.** Effect of prestress on plane shear waves in the QLV Prony series material. (a) Dimensionless dispersion relation. $\lambda$, stretch ratio, $\theta$, angle of the wave propagation direction. (b) Angular distribution of normalized phase velocity. $\lambda = 1.25$. (c) Dimensionless dissipation factor.

### 3.2 Surface and fluid-solid interface waves

For the fluid-solid interface wave, we consider a viscoelastic solid that occupies the region $x_2 \leq 0$, while the other half-space ($x_2 > 0$) is filled with an inviscid fluid. The interface wave propagates along the $x_1$ direction. Therefore, the stream function for interface waves in the solid takes the form: $\psi = \psi_0 \exp(skx_2)\exp[i(kx_1 - \omega t)]$, where $s$ is a dimensionless parameter. Inserting $\psi$ into wave motion Eq. (32), we get

$$\left(G\gamma - \Omega\sigma_{D22}^e\right)s^4 + \left[\rho\frac{\omega^2}{k^2} - 2G\beta + \Omega\left(\sigma_{D11}^e + \sigma_{D22}^e\right)\right]s^2 + G\alpha - \Omega\sigma_{D11}^e - \rho\frac{\omega^2}{k^2} = 0. \quad (41)$$

Eq. (41) is a complex quartic equation with respect to $s$, which mathematically yields four complex roots, denoted as $\pm s_1$ and $\pm s_2$. Without loss of generality, we assume that the real parts of $s_1$ and $s_2$ are nonnegative. Therefore, $\psi$ can be generally expressed as

$$\psi = [A_1 \exp(s_1 k x_2) + A_2 \exp(-s_1 k x_2) \\ + A_3 \exp(s_2 k x_2) + A_4 \exp(-s_2 k x_2)] \exp[i(k x_1 - \omega t)], \quad (42)$$

where $A_1 \sim A_4$ denote the amplitudes. To satisfy the boundedness condition of the stream function, i.e. $\psi \to 0$ as $x_2 \to -\infty$, the stream function is further simplified as follows:

$$\psi = \left[ A_1 \exp(s_1 k x_2) + A_3 \exp(s_2 k x_2) \right] \exp\left[ i(k x_1 - \omega t) \right]. \quad (43)$$

The fluid is modeled as an acoustic medium and the motion equation is

$$\frac{1}{c_p^2} \frac{\partial^2 \boldsymbol{u}^f}{\partial t^2} = \nabla^2 \boldsymbol{u}^f, \quad (44)$$

where $\boldsymbol{u}^f$ is the displacement of the fluid. $c_p = \sqrt{\kappa / \rho^f}$ is the sound speed. $\kappa$ and $\rho^f$ are bulk modulus and density of the fluid, respectively. Since $\boldsymbol{u}^f$ is an irrotational vector field, we introduce a potential function $\varphi(x_1, x_2, t)$ to replace $\boldsymbol{u}^f$ with $u_1^f = \varphi_{,1}$ and $u_2^f = \varphi_{,2}$. Inserting $\varphi$ into Eq. (44) we get

$$\varphi = B_1 \exp(-\xi k x_2) \exp\left[ i(k x_1 - \omega t) \right], \quad (45)$$

where $\xi = \sqrt{1 - \frac{\omega^2}{k^2} \frac{1}{c_p^2}}$, and $B_1$ denotes the amplitude. In Eq. (45), we have omitted the term of $\exp(\xi k x_2)$ due the boundedness condition ($\varphi \to 0$ as $x_2 \to +\infty$).

The interfacial conditions between the solid and the fluid include the continuity of normal displacement, the continuity of normal stress, and the free shear stress. These conditions can be written as follows (Li et al., 2017b; Otténio et al., 2007):

$$u_2 = u_2^f, \quad \Sigma_{21} = -\sigma_{22} u_{2,1}, \quad \Sigma_{22,1} = -p^f_{,1} - \sigma_{22} u_{2,12}, \text{ at } x_2 = 0, \quad (46)$$

where $p^f$ is the hydrostatic pressure of the fluid. Applying interfacial conditions Eq. (46), we get the secular equation for fluid-solid interface wave (i.e. Scholte wave, see derivation in Supplementary Note 5.1)

$$\begin{aligned}&\left(1+s_2^2\right)\cdot\left(-\rho\frac{\omega^2}{k^2}s_1+C_1s_1-C_2s_1^3\right)\\&-\left(1+s_1^2\right)\cdot\left(-\rho\frac{\omega^2}{k^2}s_2+C_1s_2-C_2s_2^3\right)+\left(s_1^2-s_2^2\right)\frac{\rho^f}{\xi}\frac{\omega^2}{k^2}=0\end{aligned}, \quad (47)$$

where parameters $C_1$ and $C_2$ are defined by

$$C_1 = G(2\beta+\gamma) - \Omega\left(\sigma_{D11}^e + 2\sigma_{D22}^e\right), \quad (48a)$$

$$C_2 = G\gamma - \Omega\sigma_{D22}^e, \quad (48b)$$

For the surface wave (Rayleigh wave), the stress-free boundary condition at the solid surface ($x_2 = 0$) must be satisfied, leading to the secular equation (see details in Supplementary Note 5.2)

$$\begin{aligned}&\left(1+s_2^2\right)\cdot\left(-\rho\frac{\omega^2}{k^2}s_1+C_1s_1-C_2s_1^3\right)\\&-\left(1+s_1^2\right)\cdot\left(-\rho\frac{\omega^2}{k^2}s_2+C_1s_2-C_2s_2^3\right)=0\end{aligned}. \quad (49)$$

When $\rho^f = 0$, Eq. (47) reduces to Eq. (49). When $G=1$, $\Omega=0$, elastic solutions are recovered, e.g. Eq. (47) recovers to the solution of Otténio et al. (2007); Eq. (49) recovers to the solution of Dowaikh and Ogden (1990).

The effects of prestress and material viscoelasticity on Rayleigh surface waves are next examined. It is well-know that the speed of Rayleigh surface waves vanishes when the compressive strain reaches Biot's critical strain (~0.46 under plane strain conditions), which was believed to be the onset condition for surface wrinkling on a free surface (Biot, 1963). However, both theoretical and experimental studies have revealed that another surface instability—crease—can occur prior to reaching Biot's strain (Hong et al., 2009), due to the high sensitivity of wrinkling to tiny imperfections (Cao & Hutchinson, 2012). Here we investigate the Rayleigh surface waves in a viscoelastic neo-Hookean material subjected to Biot's compressive strain. The explicit formulation of surface waves is provided in Supplementary Note 5.3. As shown in Fig.

4a, the phase velocity approaches zero when $\tilde{f} \to 0$, consistent with the onset of surface wrinkles. However, as frequency increases, the reemergence of surface wave propagation reflects frequency-dependent stiffening (i.e. material viscoelasticity) that suppresses wrinkle formation. At high frequencies, the phase velocity even exceeds that in the stress-free case, which is similar to the observation for shear waves shown in Fig. 2a. Figure 4b presents the dissipation factors. The compressive strain dramatically enlarges the dissipation even in the low frequency regime ($\tilde{f} < 10^{-4}$). For illustration, we plot the normalized wave profiles at $\tilde{f} = 10^{-4}$ for the two cases, as depicted in Fig. 4c. The nontrivially high dissipation at the Biot's strain may indicate the global wrinkling is unlikely developed on free surface, but instead, local surface instability such as crease could emerge. The implications of this result for surface morphology deserves further study, which, however, is beyond the scope of the present study.

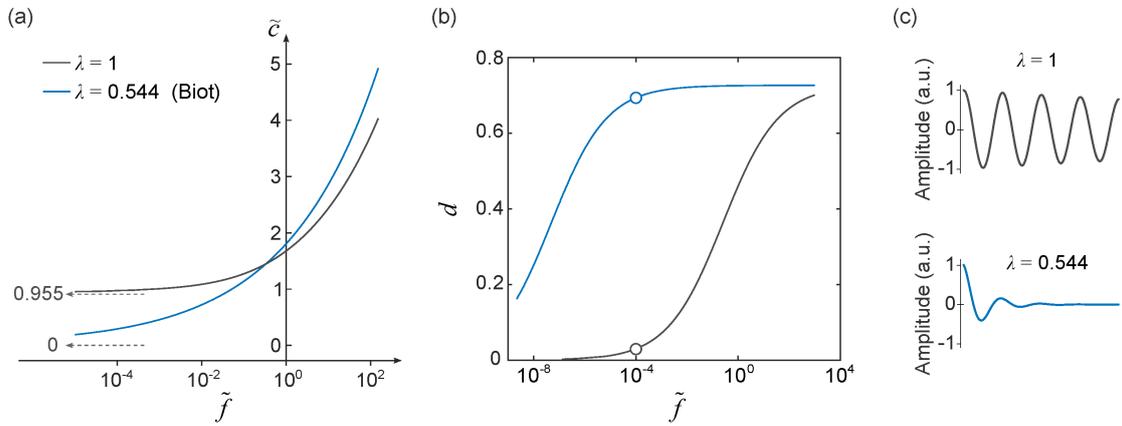

**Figure 4.** Effect of prestress on the Rayleigh surface wave in the KVFD material ($\beta_0 = 0.4$). (a) Dimensionless phase velocity $\tilde{c}$ and (b) dissipation factor $d$ of the Rayleigh surface wave at the stress-free and the compressive ($\lambda = 0.544$, Biot's strain) states. (c) Wave profiles of the Rayleigh surface waves at $\tilde{f} = 10^{-4}$ for stress-free and compressive states.

### 3.3 Lamb waves

We proceed to derive Lamb waves, which are commonly involved in guided wave elastography of thin-wall biological tissues such as arteries and corneas. We consider a prestressed viscoelastic plate immersed in inviscid fluid with a wall thickness of $2h$. The upper and lower boundary of the plate are at $x_2 = h$ and $-h$, respectively. The fluid occupies the upper ($x_2 > h$) and lower ($x_2 < -h$) space of the plate. Similarly to Sec 3.2, we introduce the stream function $\psi(x_1, x_2, t)$ that is related to the plate displacements by $u_1 = \psi_{,2}$ and $u_2 = -\psi_{,1}$. Taking $\psi$ into the wave motion equation (32), we obtain the general form of $\psi$ as follows:

$$\psi = \left[ A_1 \cosh(s_1 k x_2) + A_2 \sinh(s_1 k x_2) \right. \\ \left. + A_3 \cosh(s_2 k x_2) + A_4 \sinh(s_2 k x_2) \right] \exp\left[ i(k x_1 - \omega t) \right], \quad (50)$$

where $s_1$ and $s_2$ are the two roots of Eq. (41) with nonnegative real parts. $A_1 \sim A_4$ denote the amplitudes. It should be noted that Eq. (50) is equivalent to Eq. (42), but is presented in this form to facilitate the decomposition into symmetric and antisymmetric components. The amplitudes $A_2$ and $A_4$ vanish for antisymmetric modes, while $A_1$ and $A_3$ vanish for symmetric modes. The fluids are modeled as acoustic media. The potential functions for the upper and lower fluid are $\varphi^+ = B_1 \exp(-\xi k x_2) \exp\left[ i(k x_1 - \omega t) \right]$ and $\varphi^- = B_2 \exp(\xi k x_2) \exp\left[ i(k x_1 - \omega t) \right]$, respectively, where $\xi$ has been defined in Sec 3.2, $B_1$ and $B_2$ denotes the amplitudes.

The surfaces of the plate in contact with fluid should satisfy the continuity of normal displacement and normal stress, as well as the free shear stress. These interfacial conditions can still be expressed by Eq. (46), with the spatial position adjusted to $x_2 = \pm h$. By inserting the stream and potential functions into the interfacial conditions, the secular equation of the antisymmetric modes is (see details in Supplementary Note 6.1)

$$\left(1+s_2^2\right) \cdot \left(-\rho\frac{\omega^2}{k^2}s_1 + C_1 s_1 - C_2 s_1^3\right) \cdot \tanh(s_1 kh)$$
$$-\left(1+s_1^2\right) \cdot \left(-\rho\frac{\omega^2}{k^2}s_2 + C_1 s_2 - C_2 s_2^3\right) \cdot \tanh(s_2 kh) + \left(s_1^2 - s_2^2\right)\frac{\rho^f}{\xi}\frac{\omega^2}{k^2} = 0 \quad (51)$$

For the symmetric modes, the secular equation reads

$$\left(1+s_2^2\right) \cdot \left(-\rho\frac{\omega^2}{k^2}s_1 + C_1 s_1 - C_2 s_1^3\right) \cdot \coth(s_1 kh)$$
$$-\left(1+s_1^2\right) \cdot \left(-\rho\frac{\omega^2}{k^2}s_2 + C_1 s_2 - C_2 s_2^3\right) \cdot \coth(s_2 kh) + \left(s_1^2 - s_2^2\right)\frac{\rho^f}{\xi}\frac{\omega^2}{k^2} = 0 \quad (52)$$

where the coefficients $C_1$ and $C_2$ have been defined in Eqs. (48a) ~ (48b).

For Lamb waves in a plate in vacuum, by applying the stress-free boundary conditions at the upper and lower surfaces of the plate, the secular equation for the antisymmetric modes can be obtained as (see details in Supplementary Note 6.2)

$$\left(1+s_2^2\right) \cdot \left(-\rho\frac{\omega^2}{k^2}s_1 + C_1 s_1 - C_2 s_1^3\right) \cdot \tanh(s_1 kh)$$
$$-\left(1+s_1^2\right) \cdot \left(-\rho\frac{\omega^2}{k^2}s_2 + C_1 s_2 - C_2 s_2^3\right) \cdot \tanh(s_2 kh) = 0 \quad (53)$$

and the secular equation for the symmetric modes is

$$\left(1+s_2^2\right) \cdot \left(-\rho\frac{\omega^2}{k^2}s_1 + C_1 s_1 - C_2 s_1^3\right) \cdot \coth(s_1 kh)$$
$$-\left(1+s_1^2\right) \cdot \left(-\rho\frac{\omega^2}{k^2}s_2 + C_1 s_2 - C_2 s_2^3\right) \cdot \coth(s_2 kh) = 0 \quad (54)$$

When $\rho^f = 0$, Eqs. (51) and (52) reduce to Eqs. (53) and (54), respectively. When $G=1$, $\Omega=0$, the dispersion equations for elastic materials are recovered, e.g. Eqs. (51) and (52) recover to the solutions of Li et al. (2017b); Eqs. (53) and (54) recover to the solutions of Ogden and Roxburgh (1993). When the prestress is in absence ($\lambda_1 = \lambda_2 = \lambda_3 = 1$), the dispersion equations for linear viscoelastic materials are recovered, e.g. Eqs. (53) and (54) recover to the forms given by Rose (2014).

Figure 5 presents the dispersion and attenuation of the first four modes of Lamb waves (i.e., A0, S0, A1 and S1) in vacuum. Here we employ the neo-Hookean model and the KVFD model to describe material nonlinear viscoelasticity. The corresponding secular equations are simplified from Eqs. (53) and (54), with their explicit forms provided in Supplementary Note 6.3. We define the dimensionless frequency, phase velocity, and attenuation as $\bar{f} \equiv 2fh_0/c_t$, $\tilde{c} \equiv c/c_t$, and $\tilde{k}_{\mathrm{im}} \equiv \mathrm{Im}(k)c_t/\eta^{1/\beta_0}$, respectively, where $c_t = \sqrt{\mu/\rho}$, $h_0$ denotes the half-wall thickness in the initial (stress-free) state. Dash-dot and solid lines represent the case of $\lambda=1$ and $\lambda=1.4$ respectively. Basically, the prestress dramatically changes the phase velocity and attenuation. For the fundamental modes (A0 and S0), the tensile stress increases the phase velocity and decreases the attenuation, in line with the results for plane shear and surface waves. We find the phase velocity of A0 mode at $\bar{f} \to 0$ is $c=\sqrt{\sigma_{11}/\rho}$, where $\sigma_{11}=(G-\Omega)(\alpha-\gamma)$ is the prestress applied to the plate. This observation is consistent with elastic theory (Li et al., 2022a), indicating the prestress can be derived from dynamic responses at ultra-low frequency regime. As frequency increases, the phase velocities of the A0 and S0 modes get close to that of the Rayleigh surface wave, of which the dispersion is purely determined by viscoelasticity of the material.

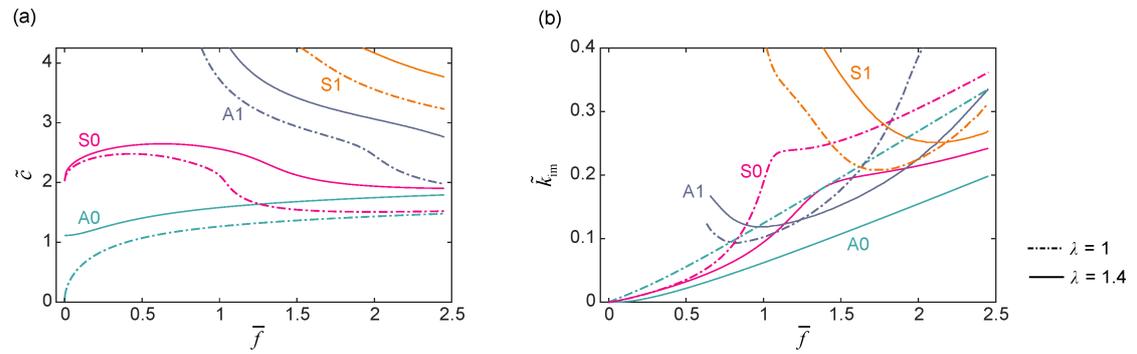

**Figure 5.** Influence of prestress on Lamb waves for the first four modes, i.e. A0, A1, S0 and S1 modes. (a) Dimensionless phase velocity $\tilde{c}$ and (b) dimensionless attenuation $\tilde{k}_{\mathrm{im}}$ ( $=k_{\mathrm{im}}c_t/\eta^{1/\beta_0}$, where $c_t=\sqrt{\mu/\rho}$ ) with respect to dimensionless frequency $\bar{f}$ ( $=2fh_0/c_t$ ). The neo-Hooke and KVFD models are used, where $\mu=40\,\mathrm{kPa}$, $\eta=0.015\,\mathrm{s}^{\beta_0}$, $\beta_0=0.4$. The wall thickness of the plate in the stress-free state is 1 mm. The plate is subject to a uniaxial stretch $\lambda_1=\lambda$, $\lambda_2=\lambda_3=\lambda^{-1/2}$.

## 4 Applications of the theory to SWE experiments

In this section, the proposed theory is applied to analyze real data obtained in experiments. We performed SWE on soft artificial materials (hydrogel and polydimethylsiloxane, PDMS) and *ex vivo* soft biological tissues (a segment of porcine ascending aorta), where surface waves or guided elastic waves (Lamb waves) were excited and detected. Harmonic stimuli over a broad frequency band or an impulse stimulus are utilized to measure the dispersion relations.

### 4.1 Optical coherence elastography of soft materials

The optical coherence elastography (OCE) system is based on a home-built swept-source optical coherence tomography (SS-OCT) platform with an A-line rate of 43.2 kHz. To perform OCE, we relied on a vibrating contact probe driven by a PZT that works in synchronization with the swept source laser. The probe generates harmonic waves with amplitudes on the order of tens of nanometers in the sample, which are detected by analyzing the phase variations of the interference signals. We used a home-built stretcher to introduce a uniaxial stretch to the sample. The wave profile on the free surface along the stretch direction was then measured, followed by a Fourier transform to extract the wavelength (Li et al., 2022a). In this way, the phase velocities at different frequencies can be obtained. More details about the experimental setup can be found in our previous work (Li et al., 2022a).

The experiments were performed on a piece of hydrogel membrane and a piece of PDMS membrane. The hydrogel sample was obtained following the protocol described in Kim et al. (2021). The thickness of the sample is about 3 mm. The PDMS was prepared by using a 2:1 mixing ratio of base elastomer and curing agent (Sylgard 184, Dow Corning) and cured at room temperature over a night. The wall thickness of the PDMS membrane is 0.47 ± 0.01 mm.

Figure 6 presents the experimental results. For the hydrogel sample, the phase velocities remain nearly constant (with variation less than 3%) across the frequency range of 8 to 20 kHz when the sample is subjected to stretch ratios of $\lambda = 1$ and $\lambda = 1.2$. These flat dispersion relations are primarily attributed to the high stimulus frequency used in the experiments, which ensured the generation of Rayleigh surface waves, and more importantly, to the extraordinary elasticity of the sample. The phase velocity increases about 21% (7.88 ± 0.05 m/s to 9.56 ± 0.09 m/s), in quantitative agreement with the stretch ratio applied to the sample. This observation indicates the superior elastic properties of the sample as our theoretical analysis suggests the effect of prestress on phase velocities will be quenched if material viscosity become significant.

For the PDMS membrane, the phase velocity shows a dramatic dispersion because the A0 mode Lamb wave is dominant. As shown in Fig. 6b-i, we find that the elastic model does not fit all the experimental data well. The fitting curve gradually deviates with the experimental data as the frequency increases. On the other hand, the KVFD viscoelastic model fits all the data well, with fitting parameters: shear modulus $\mu = 1.2$ MPa, $\eta = 0.0005\,\mathrm{s}^{\beta_0}$, and $\beta_0 = 0.65$. For the prestressed cases shown in Fig. 6b-ii, we applied the fitted material parameters into Lamb wave model with the neo-Hookean material (see Eq. (S63) in Supplementary Note 6.3) to predict wave dispersion. The theoretical predictions show good agreements with the experimental data, validating the proposed theory for modeling wave motion in prestressed viscoelastic soft materials.

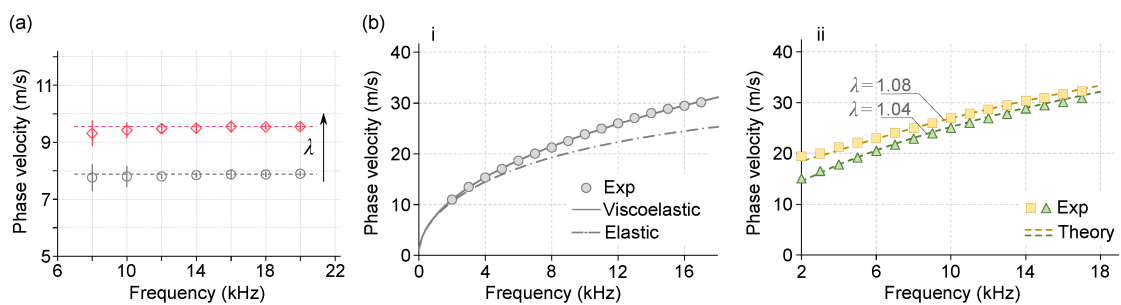

**Figure 6**. Optical coherence elastography experiments of soft materials. (a) Dispersion relations of the Rayleigh surface waves in a hydrogel. Lower markers, $\lambda = 1$. Upper

markers, $\lambda = 1.2$. The dashed lines indicate the wave velocities at 20 kHz. (b) Dispersion relations of the Lamb waves in a PDMS membrane. (i) Dispersion relations obtained at stress-free state. Markers, experiments. Solid line, fitting curve with the KVFD model. Dashed line, fitting curve with elastic model. (ii) Experimental dispersion relations (Markers) and the comparisons with the model-predicted curves (Dashed lines).

## 4.2 Ultrasound elastography of *ex vivo* soft biological tissues

The ultrasound SWE experiments were performed using the Verasonics Vantage 64LE System (Verasonics Inc., Kirkland, WA, USA), equipped with a L9-4 (central frequency 7MHz) linear array transducer (Jiarui Electronics, Shenzhen, China). The system can send long ultrasound pulses (~ 200 μs) and focus the ultrasound beam to generate a local body force (i.e., acoustic radiation force, ARF). The ARF excites elastic waves with micrometer-scale amplitudes. Then the transducer is switched to perform ultrafast plane wave imaging with a frame rate of 10 kHz, which enables the measurement of wave propagation within the imaging plane. More details of the ultrasound SWE system can be found in our previous paper (Li et al., 2022b; Zhang et al., 2023).

A segment of porcine ascending aorta was obtained from a freshly slaughtered animal. As shown in Fig. 7a, the aorta was cut off and flattened along its circumferential direction. We clamped the sample with a customized stretcher and then put the stretcher in water. The ultrasound probe (immersed in water) was hung about ~20 mm above the sample, with the imaging plane in parallel to the circumferential direction (Fig. 7b). In the imaging plane, $x_1$ and $x_2$ axes were coaxial with the circumferential and radial direction, respectively. The experiments were performed at the room temperature of 20 ℃. Figure 7c shows the spatiotemporal data acquired when the sample is subjected to different stretch, i.e., $\lambda = 1$, $1.15$, and $1.25$, respectively. By performing two-dimensional Fourier transformation to the spatiotemporal data, we obtained the dispersion relations (Figs. 7d and e). The first-order antisymmetric mode of Lamb waves (A0 mode) was primarily excited by the ARF, which is in line with previous studies (Bernal et al., 2011; Couade et al., 2010; Li et al., 2017a). The fluctuations in the experimental data likely stem from the dispersion extraction algorithm (Kijanka et al., 2019).

To analyze the data, we utilized the Gasser-Ogden-Holzapfel (GOH) model to describe arterial hyperelasticity (Gasser et al., 2006), and the KVFD model to describe arterial viscoelasticity. The strain energy function of the GOH model is

$$W = \frac{\mu}{2}(I_1 - 3) + \frac{k_1}{2k_2} \sum_{i=4,6} \left\{ \exp\left[ k_2 \left( \kappa I_1 + (1 - 3\kappa) I_i - 1 \right)^2 \right] - 1 \right\}, \tag{55}$$

where $\mu$ and $k_1$ are the initial shear modulus of elastin and collagen fibers, respectively. $k_2$ denotes the nonlinear stiffening of collagen fibers. $\kappa$ represents the fiber dispersion ($0 \leq \kappa \leq 1/3$). Invariants $I_1 = \mathrm{tr}(\boldsymbol{C})$, $I_4 = \boldsymbol{M} \cdot \boldsymbol{CM}$ and $I_6 = \boldsymbol{M}' \cdot \boldsymbol{CM}'$. $\boldsymbol{M}$ and $\boldsymbol{M}'$ denote two symmetrically distributed fiber orientations. $\phi$ denotes the angle between the fiber orientation and the circumferential direction. To get the constitutive parameters, we performed a quasi-static uniaxial tensile test to the sample, and the best-fit values are $\mu = 33.4\ \mathrm{kPa}$, $k_1 = 72.7\ \mathrm{kPa}$, $k_2 = 6.3$, $\kappa = 0.26$, $\phi = 42.8°$ (see details in Supplementary Note 8).

We then fitted the dispersion data in the stress-free state ($\lambda = 1$, solid line in Fig. 7d). The optimization function is defined by the root-mean-square error (RMSE), i.e. $\mathrm{RMSE} = \sqrt{\sum_{i=1}^{n} \left( c_i^{(\mathrm{theo})} - c_i^{(\mathrm{exp})} \right)^2 / n}$, where $c_i^{(\mathrm{theo})}$ denotes the theoretically predicted phase velocity (Eq. (S65) in Supplementary Note 6.4, with $\lambda_1 = \lambda_2 = \lambda_3 = 1$), $c_i^{(\mathrm{exp})}$ is the experimentally measured phase velocity, $n$ ( = 100) represents the number of discrete data points. The optimization process was achieved by the genetic algorithm. As a result, the viscoelastic parameters of the KVFD model are obtained as $\eta = 0.098\ \mathrm{s}^{\beta_0}$ and $\beta_0 = 0.35$. For comparison, we plot the dispersion relation predicted by the elastic model (i.e., inserting $\eta = 0$ into Eq. (S65)). The elastic curve is reasonably lower than the experimental data as the constitutive parameters are obtained from quasi-static tests, whereas the central frequency of the Lamb waves is about 500 Hz.

With all the fitting parameters from the tensile test and wave dispersion in the stress-free state, we predict the dispersion relations when the sample is subjected to prestress ($\lambda = 1.15$ and $1.25$, using Eq. (S65)), as shown in Fig. 7e. The model predictions are in excellent agreement with the experimental data (relative error < 1.5% over 500 Hz), validating the effectiveness of the proposed theory in modeling wave propagation in biological tissues exhibiting both strong nonlinear elasticity and significant viscosity.

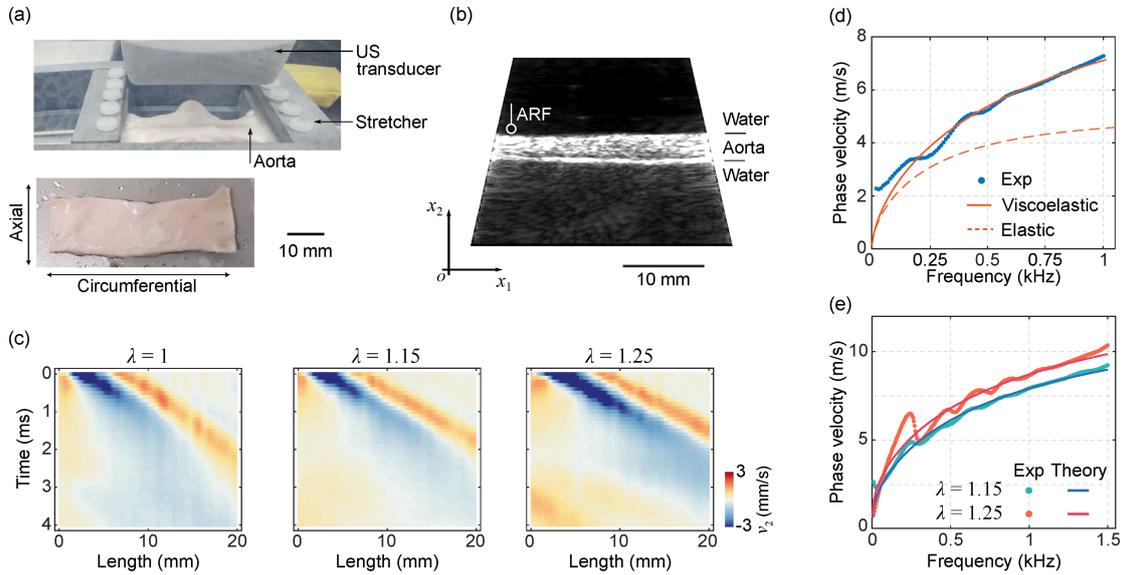

**Figure 7**. Ultrasound elastography experiments on an ex vivo porcine ascending aorta. (a) Experimental setup (top), and photography of the sample (bottom). (b) Ultrasound B-mode image of the sample. Circumferential direction is parallel to the imaging plane. ARF, acoustic radiation force. (c) Spatiotemporal maps of particle velocity showing guided elastic wave propagations in the sample when subject to different prestretch. From left to right, $\lambda = 1$, $1.15$, $1.25$. (d) Experimental dispersion relation obtained in stress-free state ($\lambda = 1$). Solid line: fitting curve using the viscoelastic model. Dashed line, theoretical curve with the elastic model. The constitutive parameters were obtained by tensile test (see supplementary Note 8). (e) Experimental dispersion relations (Markers) and the comparisons with the viscoelastic model-predicted curves (Solid lines).

## 5  Acoustoelastic imaging of stresses in viscoelastic solids

We proceed to study the measurement of stress in viscoelastic solids with small-amplitude elastic waves. We firstly give a relationship between the relaxed stresses and incremental parameters that is free from constitutive model. Based on this principle, we reveal that the squared difference of complex shear wave velocities along the two principle axes is related to the difference in relaxed stresses along the corresponding directions, which leads to a promising method for stress measurement in viscoelastic materials. Finally, we validate the proposed method using finite element analysis.

### 5.1  Relationship between principal stresses and incremental parameters

The relaxed stress at the deformed configuration ($\mathcal{B}$) can be expressed in a general form as

$$\boldsymbol{\sigma} = -q\boldsymbol{I} + (G - \Omega)\boldsymbol{\sigma}_D^e. \tag{56}$$

Combining Eqs. (34), (35) and (56), the difference between the normal stresses along the $x_1$ and $x_2$ directions, denoted as $\sigma_{11}$ and $\sigma_{22}$, respectively, is obtained as

$$\sigma_{11} - \sigma_{22} = (G - \Omega)(\alpha - \gamma). \tag{57}$$

For the Prony series model, $G - \Omega = 1 - g$; $\alpha$ and $\gamma$ denote the instantaneous incremental parameters. For the KVFD model, $G - \Omega = 1$; $\alpha$ and $\gamma$ denote the long-term parameters. In general, Eq. (57) suggests the difference of the principal stresses equals to the difference of the long-term incremental parameters. Note that, since the incremental parameters $\alpha$ and $\gamma$ are directly related to the multiple elastic waves presented in Sec. 3, Eq. (57) suggests a potential approach to characterize internal prestress by extracting $\alpha$ and $\gamma$ from wave measurements.

### 5.2  Measurement of stresses by plane shear waves

We proceed to examine a specific type of waves—plane shear waves—to demonstrate how stress can be characterized through wave measurements. For plane shear waves, we denote the complex wave velocity along the principal directions $x_1$ and $x_2$ as $\mathcal{C}_1$ and $\mathcal{C}_2$, respectively. According to Eq. (36), we get

$$\rho \mathcal{C}_1^2 = G\alpha - \Omega \sigma_{D11}^e, \tag{58}$$

and

$$\rho \mathcal{C}_2^2 = G\gamma - \Omega \sigma_{D22}^e. \tag{59}$$

With the help of Eqs. (34) and (35), it can be deduced that

$$\rho \mathcal{C}_1^2 - \rho \mathcal{C}_2^2 = (G - \Omega)(\alpha - \gamma). \tag{60}$$

Comparing Eqs. (60) and (57), we finally obtain

$$\rho \mathcal{C}_1^2 - \rho \mathcal{C}_2^2 = \sigma_{11} - \sigma_{22}. \tag{61}$$

Eq. (61) represents an extension of our previous work based on purely elastic models (Li et al., 2022a; Zhang et al., 2023). It offers a promising and general approach to probe stresses via plane shear waves, applicable to various material models (including both isotropic and anisotropic hyperelastic models, as well as QLV and KVFD viscoelastic models). Moreover, Eq. (61) is frequency-independent, indicating that long-term stresses in viscoelastic materials can, in principle, be measured from plane shear waves at any given frequency ranges, depending on the elastography modality employed (Ormachea & Parker, 2020). The frequency-independency of the squared difference between $\mathcal{C}_1$ and $\mathcal{C}_2$ is somewhat surprising, given that both $\mathcal{C}_1$ and $\mathcal{C}_2$ are frequency-dependent. We will further verify this relation using finite element analysis in Sec 5.3.

To implement Eq. (61) using experimental data, the complex wave velocity of plane shear waves can be acquired by measuring the phase velocity $c$ and wave attenuation $k_{im}$ at any given frequency $\omega$. Then the real wavenumber is calculated by $k_{re} = \omega/c$, and the complex wave velocity can be obtained using $\mathcal{C} = \omega/(k_{re} + ik_{im})$.

## 5.3 Verification of the method using finite element analysis

We performed finite element analysis to verify the proposed method. The finite element model was built by Abaqus/CAE 6.14 (Dassault Systemes, USA). Figure 8a depicts the model. We built a 2D square domain and prescribed the in-plane stretch ($\lambda_1 = 2$, $\lambda_2 = 0.5$, $\lambda_3 = 1$) to introduce prestress. Then a harmonic line force was applied to generate plane shear waves. Approximately 100,000 solid elements (CPE8RH) were used to discrete the domain. Convergence of the simulation was carefully examined by refining the mesh size and time increment. Figure 8b and c show the wave profiles along the $x_1$ and $x_2$ axes, respectively, obtained at 5 kHz. The prestress results in a higher phase velocity ($c$) and lower attenuation ($k_\text{im}$) along the $x_1$ axis compared to the $x_2$ axis. Figures 8d and e present the phase velocities and attenuations at different frequencies derived from the wave profiles (methods of measuring phase velocity and attenuation are detailed in Supplementary Note 9). The results obtained from FEA match well with the theory. We then derive the stress $\sigma_{11}$ from the phase velocities and attenuations, as shown in Fig. 8f. Using the proposed method, we get a consistent value for $\sigma_{11}$ from the phase velocities and attenuations at different frequencies, which is in excellent agreement (relative error < 1.5%) with the applied prestress (150 kPa). These results validate the effectiveness of the proposed method and demonstrate a potential experimental setup for stress measurement in viscoelastic solids.

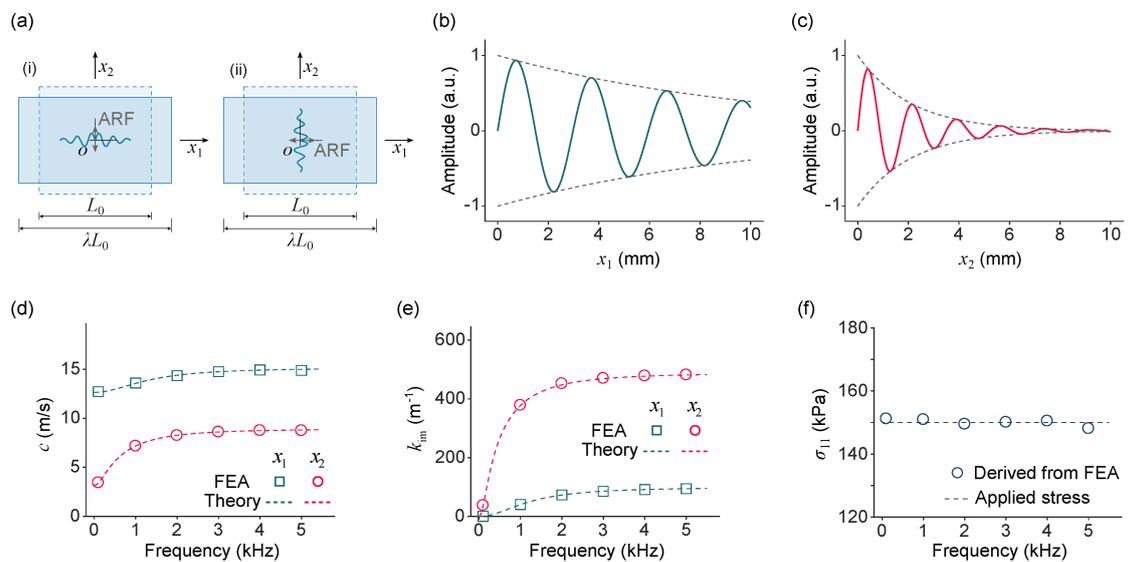

**Figure 8.** Verification of stress measurement for viscoelastic materials using finite element analysis (FEA). (a) Schematic of the model. (i) and (ii) show shear wave generation and propagation in $x_1$ and $x_2$ axes, respectively. The stretch is applied along the $x_1$ axis with a stretch ratio $\lambda$. A harmonic line load perpendicular to the wave propagation direction is applied to excite waves. (b) and (c) Wave profiles along the $x_1$ and $x_2$ axes, respectively. Dashed lines with exponential decay outline the attenuation of wave amplitudes. Frequency, 5 kHz. (d) and (e) Phase velocities and attenuations at different frequencies derived from the wave profiles. Markers, FEA. Dashed lines, theory. (f) Comparison of measured stress and applied stress. Dots, stress derived from the phase velocity and attenuation at each excitation frequency. Dashed line, applied stress ($\sigma_{11} = 150$ kPa). The material models used in this analysis are the neo-Hookean material and one-term Prony series. Parameters are $\mu = 40$ kPa (long-term), $g = 0.5$, $\tau = 0.1$ ms. The stretch ratio is $\lambda = 2$ ($\lambda_1 = \lambda$, $\lambda_2 = \lambda^{-1}$, $\lambda_3 = 1$).

## 6  Discussion and conclusions

An incremental dynamics theory for prestressed viscoelastic solids is proposed in this work, of which two viscoelastic models are considered: the QLV Prony series model and the KVFD model. Based on this theory, the analytical solution of three classes of representative elastic waves—commonly involved in SWE of soft tissues—are derived: plane shear waves, surface/fluid-solid interface waves, and Lamb waves. The key features of their dispersion and attenuation behaviors under prestress and material viscoelasticity are investigated. Interestingly, for the KVFD model, the effect of prestress on phase velocity will be quenched at high-frequency range, resulting in an isotropic wave front even in the presence of anisotropic prestress.

SWE measurements, including optical coherence elastography and ultrasound elastography, were performed on soft artificial materials and *ex vivo* porcine tissues, respectively, to validate the proposed theory. When incorporated with the KVFD model, our theory matches the experimental dispersion across a broad frequency band, which demonstrates its capability and provides a theoretical basis for characterizing both viscoelasticity and prestress effects in soft materials. It should be noted that the analytical solutions of multiple elastic waves derived in this work are applicable to arbitrary hyperelastic constitutive models. Therefore, the present work enables characterizing multiple elastic wave propagation in biological soft tissues, particularly when considering their fiber-reinforcing features through constitutive models such as the Demiray-Fung model (Demiray, 1972) and the GOH model (Gasser et al., 2006). This makes the proposed theory especially relevant and valuable for the mechanical characterization of biological soft tissues.

Based on the theory, we further reveal that the static relaxed prestress in a viscoelastic solid can be readily determined from plane shear wave motions, independent of wave frequency. This finding leads to an approach to measure prestress via phase velocities and attenuations of plane shear waves propagating along mutually orthogonal principal directions, without prior knowledge of constitutive parameters and applicable across broad measurement frequencies. This is an extension of the conclusion for purely elastic material (Li et al., 2022a; Zhang et al., 2023).

The viscoelastic models (both the QLV and KVFD model) adopted in this work assume that the viscous stress depends solely on the stress history. Although this assumption is simple within the nonlinear viscoelastic field, our SWE experiments demonstrate its validity. A recent study proposed strain-rate-dependent fractional derivative viscoelastic models and presented corresponding solutions for plane shear waves (Berjamin & Destrade, 2025). Incorporating more general viscoelastic models into the incremental dynamics theory shows promise and warrants further investigation.

In conclusion, the incremental dynamics of prestressed viscoelastic solids presented in this study shall find applications in future developments of spatially resolved SWE techniques, and more broadly, provides insight into wave motions in soft materials.

**Declaration of competing interest**

The authors declare that they have no known competing financial interests or personal relationships that could have appeared to influence the work reported in this paper.

**Supplementary materials**

The supplementary materials are provided in a document file.

# Supplementary Materials

# Incremental dynamics of prestressed viscoelastic solids and its applications in shear wave elastography

Yuxuan Jiang[1], Guo-Yang Li[2,*], Zhaoyi Zhang[1], Shiyu Ma[1], Yanping Cao[1,*], Seok-Hyun Yun[3,4]

[1] Institute of Biomechanics and Medical Engineering, AML, Department of Engineering Mechanics, Tsinghua University, Beijing 100084, PR China

[2] Department of Mechanics and Engineering Science, College of Engineering, Peking University, Beijing 100871, PR China

[3] Harvard Medical School and Wellman Center for Photomedicine, Massachusetts General Hospital, Boston, Massachusetts 02114, USA

[4] Harvard-MIT Division of Health Sciences and Technology, Cambridge, MA 02139, USA

---

[*]Corresponding author: Guo-Yang Li (lgy@pku.edu.cn); Yanping Cao (caoyanping@tsinghua.edu.cn).

## Supplementary Note 1. Index notation and conventions

The first convention is about the divergence: $\text{div}(*) = \boldsymbol{e}_i \cdot \frac{\partial(*)}{\partial x_i}$. Accordingly, the divergence of a second-order tensor field $\mathbf{A}$ is expressed as $\text{div}(\mathbf{A}) = A_{ij,i}$, where indices after the coma denote spatial differentiation, and summation over repeated indices is performed. This definition is consistent with some work (Destrade, 2015; Destrade et al., 2009; Ogden, 1997; Ogden, 2003; Ogden, 2007). Based on the above definition for divergence, the equilibrium equations are expressed in terms of nominal stress ($\boldsymbol{S}$):

$$\text{Div}\boldsymbol{S} = \rho \boldsymbol{x}_{,tt}, \tag{S1}$$

with its component form

$$\frac{\partial S_{\alpha i}}{\partial X_\alpha} = \rho x_{i,tt}. \tag{S2}$$

Eq. (S1) has been adopted in Section 2.1.2. In some other texts (Berjamin & De Pascalis, 2022; Holzapfel, 2002), another definition of divergence was adopted: $\text{div}(\mathbf{A}) = A_{ij,j}$. In that case, the equilibrium equations should be formulated in terms of the first Piola-Kirchhoff stress.

The second convention is about the ordering of the indices in the partial derivative with respect to the deformation gradient ($\boldsymbol{F}$): the derivative of $\mathbf{A} = \frac{\partial(*)}{\partial \boldsymbol{F}}$ is written as $A_{\alpha i} = \frac{\partial(*)}{\partial F_{i\alpha}}$, where $(*)$ is a scalar. This definition is consistent with the previously referenced work (Destrade, 2015; Ogden, 2003; Ogden, 2007). With this convention, the elastic stress-deformation relations can be written as:

$$\boldsymbol{\sigma} = -p\mathbf{I} + \boldsymbol{F}\frac{\partial W}{\partial \boldsymbol{F}} \tag{S3}$$

for the Cauchy stress, and

$$\boldsymbol{T} = -p\mathbf{C}^{-1} + \frac{\partial W}{\partial \boldsymbol{F}}\boldsymbol{F}^{-\text{T}} \tag{S4}$$

for the second Piola-Kirchhoff stress. $W$ denotes the strain energy function. $p$ is a Lagrangian multiplier for incompressible materials. Eqs. (S3) and (S4) have been adopted in Eqs. (23) and (24). In some other texts (Holzapfel, 2002), the ordering of the indices is $A_{i\alpha} = \frac{\partial (*)}{\partial F_{i\alpha}}$. In that case, the stress-deformation relations become (taking the Cauchy stress as an example): $\boldsymbol{\sigma} = -p\mathbf{I} + \boldsymbol{F}\left(\frac{\partial W}{\partial \boldsymbol{F}}\right)^{\mathrm{T}}$.

**Supplementary Note 2. Verification of theoretical dispersion relations using finite element analysis**

In order to verify theoretical solutions of wave dispersion and attenuation, a finite element analysis (FEA) was performed using Abaqus/CAE 6.14 (Dassault Systemes, USA). As shown in Fig. S1a, a two-dimensional finite element model was built to verify theoretical plane shear waves. We adopted incompressible neo-Hooke constitutive model to describe material hyperelasticity and the one-term Prony series to describe material viscoelasticity. Firstly, the bulk material was pre-stretched along $x_1$ with the stretch ratios $\lambda_1 = 2$ and $\lambda_3 = 1$. Then a body force with a spatial Gaussian distribution and a temporal sinusoidal oscillation was applied on material. This single frequency excitation varies from 100 Hz to 5 kHz in the simulation. The particle velocity field along the horizontal path in $x_1$ was extracted. The phase velocity and attenuation were measured according to the methods described in Supplementary Note 9. Similar to the finite element analysis of shear waves, the Rayleigh surface waves and Lamb waves were motivated and measured as shown in Fig. S1b and Fig. S1c, respectively. For the model of shear waves, approximately 100,000 solid elements (CPE8RH) were used to discrete the bulk material. For the model of surface waves, approximately 10,000 solid elements (CPE8RH) were used to discrete the solid layer. For the model of Lamb waves, approximately 1500 solid elements (CPE8RH) were applied to discrete the plate. Convergence of the simulation was carefully examined by comparing the computational results with those given by a refining mesh. As shown, the theoretical models and FEA results are basically consistent (relative error < 2%).

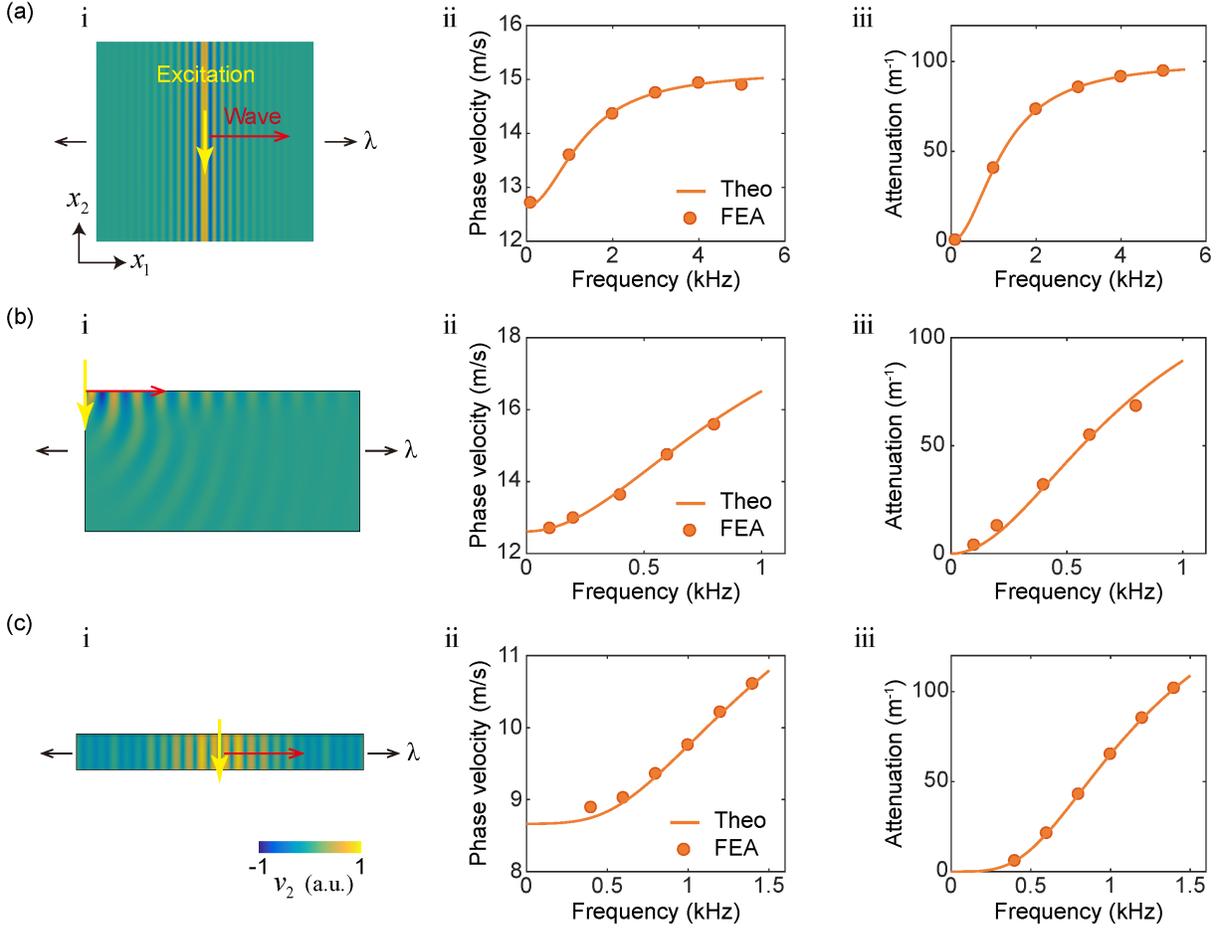

**Figure S1.** Verification of theoretical solutions by finite element analysis. **(a)** Plane shear waves, including **i**. finite element model; **ii**. dispersion curve; **iii**. attenuation curve. The material parameters include initial shear modulus $\mu = 40$ kPa (long-term), viscoelastic parameters $g = 0.5$, $\tau = 0.1$ ms, and density $\rho = 1000$ kg/m$^3$. The material is subjected to an in-plane stretch, with stretch ratios $\lambda_1 = 2$, $\lambda_2 = 0.5$, and $\lambda_3 = 1$. **(b)** Rayleigh surface waves, including **i**. finite element model; **ii**. dispersion curve; **iii**. attenuation curve. The material parameters include $\mu = 40$ kPa (long-term), $g = 0.8$, $\tau = 0.1$ ms. The solid layer is subjected to an in-plane uniaxial stretch with $\lambda_1 = 2$. **(c)** The A0 mode of Lamb waves in vacuum, including **i**. finite element model; **ii**. dispersion curve; **iii**. attenuation curve. The material parameters include $\mu = 20$ kPa (long-term), $g = 0.8$, $\tau = 0.1$ ms. The wall thickness of the plate is 4 mm before stretching. The plate is subjected to an in-plane uniaxial stretch with $\lambda_1 = 2$.

**Supplementary Note 3. Comparison of current and reported plane shear wave solutions under the Mooney-Rivlin model and specific deformation**

A previous study derived an analytical solution for plane shear waves, in which the material was modeled using the Mooney-Rivlin model for the hyperelastic part and a one-term Prony series for the viscoelastic part (Berjamin & De Pascalis, 2022). The strain energy function of the Mooney-Rivlin model is $W = C_{10}(I_1 - 3) + C_{01}(I_2 - 3)$ (see details in Supplementary Note 7.2). In their work, the wave was assumed to propagate along the $x$ direction, and the material was subject to uniaxial stretch along the $y$ direction, with deformation gradient tensor $\boldsymbol{F} = \mathrm{diag}(\lambda^{-1/2}, \lambda, \lambda^{-1/2})$. The dispersion relationship is (see Eqs. (30) and (31) therein):

$$\rho \frac{\omega^2}{k^2} = (1 - g_1)\left[\bar{\boldsymbol{T}}_d^e\right]_{11} + \left(1 - \frac{g_1}{1 + i\omega\tau_1}\right)\bar{\mu}_x^v, \tag{S5}$$

where

$$\left[\bar{\boldsymbol{T}}_d^e\right]_{11} = \frac{2}{3}\left[(3\lambda^{-1} - I_1)C_{10} + (3I_1\lambda^{-1} - 3\lambda^{-2} - 2I_2)C_{01}\right], \tag{S6}$$

$$\bar{\mu}_x^v = \frac{2}{3}\left[I_1 C_{10} + (2I_2 - 3\lambda)C_{01}\right]. \tag{S7}$$

$G$ and $\Omega$ have already been defined in Eqs. (19) and (20), their one-term forms are

$$G = 1 - \frac{g_1}{1 + i\omega\tau_1}, \quad \Omega = \frac{ig_1\omega\tau_1}{1 + i\omega\tau_1}, \tag{S8}$$

and they satisfy the identity $G - \Omega = 1 - g_1$. Using this identity, we can rewrite Eq. (S5) as follows

$$\rho \frac{\omega^2}{k^2} = G\left(\bar{\mu}_x^v + \left[\bar{\boldsymbol{T}}_d^e\right]_{11}\right) - \Omega\left[\bar{\boldsymbol{T}}_d^e\right]_{11}. \tag{S9}$$

In the case of uniaxial stretch along the $y$ direction, the invariants are $I_1 = 2\lambda^{-1} + \lambda^2$, and $I_2 = 2\lambda + \lambda^{-2}$. Inserting $I_1$ and $I_2$ into Eqs. (S6) and (S7), we get

$$\left[\bar{\boldsymbol{T}}_d^e\right]_{11} = \frac{2}{3}\left[(\lambda^{-1} - \lambda^2)C_{10} + (\lambda^{-2} - \lambda)C_{01}\right], \tag{S10}$$

$$\bar{\mu}_x^v = \frac{2}{3}\left[(2\lambda^{-1} + \lambda^2)C_{10} + (\lambda + 2\lambda^{-2})C_{01}\right]. \tag{S11}$$

Applying Eqs. (S10) - (S11) into Eq. (S9), we finally obtain

$$\rho \frac{\omega^2}{k^2} = G\left(2C_{10}\lambda^{-1} + 2C_{01}\lambda^{-2}\right) - \Omega\left[\frac{2}{3}C_{10}\left(\lambda^{-1} - \lambda^2\right) + \frac{2}{3}C_{01}\left(\lambda^{-2} - \lambda\right)\right]. \tag{S12}$$

In the following, we derive the solution of the plane shear waves based our theory. Since the wave propagates along the $x_1$ direction, we can insert $\theta = 0$ into Eq. (36) to get the plane shear wave

$$\rho \frac{\omega^2}{k^2} = G\alpha - \Omega \sigma^e_{D11}, \tag{S13}$$

where $\alpha$ and $\sigma^e_{D11}$ have been explicitly expressed by Eqs. (S74) and (S75) (see details in Supplementary Note 7.2). Combining the deformation conditions $\lambda_1 = \lambda_3 = \lambda^{-1/2}$ and $\lambda_2 = \lambda$, they are expressed as

$$\alpha = 2C_{10}\lambda^{-1} + 2C_{01}\lambda^{-2}, \tag{S14}$$

$$\sigma^e_{D11} = \frac{2}{3}C_{10}\left(\lambda^{-1} - \lambda^2\right) + \frac{2}{3}C_{01}\left(\lambda^{-2} - \lambda\right). \tag{S15}$$

Inserting Eqs. (S14) - (S15) into Eq. (S13), we again get Eq. (S12), which is the solution shown in the main text (i.e. Eq. (39)). In other words, it can be easily verified that $\bar{\mu}^v_x + \left[\bar{T}^e_d\right]_{11}$ in Eq. (S9) is equal to $\alpha$ in Eq. (S13), and $\left[\bar{T}^e_d\right]_{11}$ in Eq. (S9) is equal to $\sigma^e_{D11}$ in Eq. (S13). Therefore, Eq. (S9) is equivalent to Eq. (S13).

Thus, the plane shear wave solution reported in the literature (Berjamin & De Pascalis, 2022) is shown to be equivalent to the one derived in this study.

**Supplementary Note 4. Plane shear wave under the neo-Hookean model and specific deformation**

Here we consider a viscoelastic material subjected to uniaxial stretch along the $x_1$ direction, with in-plane deformation confined to the $x_1 - x_2$ plane. The corresponding deformation gradient tensor is $\boldsymbol{F} = \text{diag}(\lambda, \lambda^{-1}, 1)$. The hyperelastic behavior of the material is described using the neo-Hookean model. Substituting the current deformation into Eqs. (S71) and (S72) (see Supplementary Note 7.1), we obtain the explicit forms of the incremental parameters:

$$\alpha = \mu\lambda^2, \quad \gamma = \mu\lambda^{-2}, \quad \beta = \frac{\mu}{2}(\lambda^2 + \lambda^{-2}), \tag{S16}$$

and the deviatoric stresses

$$\sigma_{D11}^e = \frac{\mu}{3}(2\lambda^2 - \lambda^{-2} - 1), \quad \sigma_{D22}^e = \frac{\mu}{3}(2\lambda^{-2} - \lambda^2 - 1). \tag{S17}$$

Inserting Eqs. (S16) - (S17) into Eq. (36), we obtain the solution of plane shear waves propagating in the $x_1 - x_2$ plane:

$$\rho\mathcal{C}^2 = \left[G\mu\lambda^2 - \frac{1}{3}\Omega\mu(2\lambda^2 - \lambda^{-2} - 1)\right]\cos^4\theta + \left[G\mu\lambda^{-2} - \frac{1}{3}\Omega\mu(2\lambda^{-2} - \lambda^2 - 1)\right]\sin^4\theta$$
$$+ \left[G\mu(\lambda^2 + \lambda^{-2}) - \frac{1}{3}\Omega\mu(\lambda^2 + \lambda^{-2} - 2)\right]\sin^2\theta\cos^2\theta, \tag{S18}$$

where $\theta$ denotes the angle between the wave propagation direction and the $x_1$ axis. When the shear wave propagates along the $x_1$ direction, the complex wave velocity is

$$\rho\mathcal{C}_1^2 = G\mu\lambda^2 - \frac{1}{3}\Omega\mu(2\lambda^2 - \lambda^{-2} - 1). \tag{S19}$$

When the shear wave propagates along the $x_2$ direction, the complex wave velocity is

$$\rho\mathcal{C}_2^2 = G\mu\lambda^{-2} - \frac{1}{3}\Omega\mu(2\lambda^{-2} - \lambda^2 - 1). \tag{S20}$$

For the KVFD model, we can further simplify Eqs. (S18) – (S20) as follows:

$$\rho\mathcal{C}^2 = \left[\mu\lambda^2 + \frac{1}{3}\mu\eta(i\omega)^{\beta_0}(\lambda^2 + \lambda^{-2} + 1)\right]\cos^4\theta$$

$$+ \left[\mu\lambda^{-2} + \frac{1}{3}\mu\eta(i\omega)^{\beta_0}(\lambda^2 + \lambda^{-2} + 1)\right]\sin^4\theta \quad , \quad \text{(S21)}$$

$$+ \left[\mu(\lambda^2 + \lambda^{-2}) + \frac{2}{3}\mu\eta(i\omega)^{\beta_0}(\lambda^2 + \lambda^{-2} + 1)\right]\sin^2\theta\cos^2\theta$$

$$\rho\mathcal{C}_1^2 = \mu\lambda^2 + \frac{1}{3}\mu\eta(i\omega)^{\beta_0}(\lambda^2 + \lambda^{-2} + 1), \quad \text{(S22)}$$

and

$$\rho\mathcal{C}_2^2 = \mu\lambda^{-2} + \frac{1}{3}\mu\eta(i\omega)^{\beta_0}(\lambda^2 + \lambda^{-2} + 1), \quad \text{(S23)}$$

respectively. $\eta$ and $\beta_0$ are the two viscoelastic parameters of the KVFD model.

For the one-term Prony series model, the plane shear wave propagating in the $x_1 - x_2$ plane, along the $x_1$ direction, and along the $x_2$ direction are expressed by (note that $\mu$ should be replaced by $\mu/(1 - g_1)$ in Eqs. (S18) – (S20) to account for instantaneous modulus):

$$\rho\mathcal{C}^2 = \left[\mu\lambda^2 + \frac{\mu g_1}{3(1-g_1)}\frac{i\omega\tau_1}{1+i\omega\tau_1}(\lambda^2 + \lambda^{-2} + 1)\right]\cos^4\theta$$

$$+ \left[\mu\lambda^{-2} + \frac{\mu g_1}{3(1-g_1)}\frac{i\omega\tau_1}{1+i\omega\tau_1}(\lambda^2 + \lambda^{-2} + 1)\right]\sin^4\theta \quad , \quad \text{(S24)}$$

$$+ \left[\mu(\lambda^2 + \lambda^{-2}) + \frac{2\mu g_1}{3(1-g_1)}\frac{i\omega\tau_1}{1+i\omega\tau_1}(\lambda^2 + \lambda^{-2} + 1)\right]\sin^2\theta\cos^2\theta$$

$$\rho\mathcal{C}_1^2 = \mu\lambda^2 + \frac{\mu g_1}{3(1-g_1)}\frac{i\omega\tau_1}{1+i\omega\tau_1}(\lambda^2 + \lambda^{-2} + 1), \quad \text{(S25)}$$

and

$$\rho\mathcal{C}_2^2 = \mu\lambda^{-2} + \frac{\mu g_1}{3(1-g_1)}\frac{i\omega\tau_1}{1+i\omega\tau_1}(\lambda^2 + \lambda^{-2} + 1), \quad \text{(S26)}$$

respectively. $g_1$ and $\tau_1$ are the two viscoelastic parameters of the one-term Prony series model.

Given the complex wave velocity $\mathcal{C}$, the phase velocity $c$ can then be calculated by $c = \omega/\text{Re}(k) = \left[\text{Re}(\mathcal{C}^{-1})\right]^{-1}$.

**Supplementary Note 5. Derivation of secular equations of surface waves and fluid-solid interface waves**

**S5.1 Fluid-solid interface wave**

The stream function of the solid layer is

$$\psi = \left[A_1 \exp(s_1 k x_2) + A_3 \exp(s_2 k x_2)\right] \exp\left[i(k x_1 - \omega t)\right], \quad (S27)$$

where $s_1$ and $s_2$ are the two roots solved by Eq. (41). The potential function of the fluid is

$$\varphi = B_1 \exp(-\xi k x_2) \exp\left[i(k x_1 - \omega t)\right], \quad (S28)$$

where $\xi = \sqrt{1 - \omega^2/(k^2 c_p^2)}$. The fluid-solid interface conditions include

$$u_2 = u_2^f, \quad \Sigma_{21} = -\sigma_{22} u_{2,1}, \quad \Sigma_{22,1} = -p^f{}_{,1} - \sigma_{22} u_{2,12}, \text{ at } x_2 = 0. \quad (S29)$$

Using Eq. (26), the incremental stress is related to the stream function as follows

$$\Sigma_{21} = -\left[(G-\Omega)\sigma^e_{D22} - \sigma_{22} + G\mathcal{A}_{01221} + GQ\right]\psi_{,11} + \left(G\gamma - \Omega\sigma^e_{D22}\right)\psi_{,22}, \quad (S30)$$

$$\Sigma_{22,1} = -\left[(G-\Omega)\sigma^e_{D22} - \sigma_{22} + G\mathcal{A}_{01221} + GQ + 2G\beta - \Omega\left(\sigma^e_{D11} + \sigma^e_{D22}\right)\right]\psi_{,112}$$
$$- \left(G\gamma - \Omega\sigma^e_{D22}\right)\psi_{,222} + \rho\psi_{,2tt} \quad . \quad (S31)$$

The fluid pressure is related to the potential function as

$$p^f = -\kappa\left(\varphi_{,11} + \varphi_{,22}\right). \quad (S32)$$

Applying Eqs. (S27) – (S28) into the incremental stresses (Eqs. (S30) – (S31)), fluid pressure (Eq. (S32)) and displacements (via $u_1 = \psi_{,2}$, $u_2 = -\psi_{,1}$, $u_1^f = \varphi_{,1}$, $u_2^f = \varphi_{,2}$) to get their harmonic forms, and then substituting those harmonic forms of stresses and displacements into boundary conditions Eq. (S29), we get a system of linear equations

$$\mathbf{L}^{(\text{Scholte})}_{3\times 3}\left[A_1, A_3, B_1\right]^{\mathrm{T}} = 0, \quad (S33)$$

where the components of the matrix $\mathbf{L}^{(\text{Scholte})}_{3\times 3}$ include

$L_{11} = 1$, $L_{12} = 1$, $L_{13} = i\xi$,

$$L_{21} = 1 + s_1^2, \quad L_{22} = 1 + s_2^2, \quad L_{23} = 0,$$

$$L_{31} = -\rho \frac{\omega^2}{k^2} s_1 + C_1 s_1 - C_2 s_1^3, \quad L_{32} = -\rho \frac{\omega^2}{k^2} s_2 + C_1 s_2 - C_2 s_2^3, \quad L_{33} = i\rho^f \frac{\omega^2}{k^2}. \tag{S34}$$

where $C_1$ and $C_2$ have been defined in Eq. (48). $\rho$ and $\rho^f$ denote the material densities of solid layer and fluid, respectively. $i$ in the elements $L_{13}$ and $L_{33}$ denotes the imaginary unit. To ensure the existence of non-trivial solutions in Eq. (S33), we have

$$\det\left(\mathbf{L}_{3\times3}^{(\text{Scholte})}\right) = 0. \tag{S35}$$

By expanding Eq. (S35), the secular equation for the fluid-solid interface wave (Scholte wave) can be obtained (i.e. Eq. (47) in the main text).

### S5.2 Surface wave

The boundary conditions of the solid layer include

$$\Sigma_{21} = 0, \quad \Sigma_{22,1} = 0, \text{ at } x_2 = 0. \tag{S36}$$

Applying Eqs. (S30) – (S31) into the above boundary conditions, we obtain the following linear equations

$$\mathbf{L}_{2\times2}^{(\text{Rayleigh})} [A_1, A_3]^\text{T} = 0, \tag{S37}$$

where the components of the matrix $\mathbf{L}_{2\times2}^{(\text{Rayleigh})}$ include

$$L_{11} = 1 + s_1^2, \quad L_{12} = 1 + s_2^2,$$

$$L_{21} = -\rho \frac{\omega^2}{k^2} s_1 + C_1 s_1 - C_2 s_1^3, \quad L_{22} = -\rho \frac{\omega^2}{k^2} s_2 + C_1 s_2 - C_2 s_2^3. \tag{S38}$$

To ensure the existence of non-trivial solutions in Eq. (S37), we have

$$\det\left(\mathbf{L}_{2\times2}^{(\text{Rayleigh})}\right) = 0. \tag{S39}$$

By expanding Eq. (S39), the secular equation for the surface wave (Rayleigh wave) can be obtained (i.e. Eq. (49) in the main text, where we have assumed $\sigma_{22} = 0$).

## S5.3 Specific form of surface waves under neo-Hookean model

Here we apply a specific constitutive model—neo-Hookean model, and present the corresponding secular equation of the surface waves. The KVFD model is used to describe material viscoelasticity. Inserting explicit forms given by Eqs. (S71) and (S72) (see Supplementary Note 7.1) into Eq. (49), the secular equation can be simplified as

$$\left(1+s_2^2\right)\cdot\left(-\rho\frac{\omega^2}{k^2}s_1 + C_1 s_1 - C_2 s_1^3\right) - \left(1+s_1^2\right)\cdot\left(-\rho\frac{\omega^2}{k^2}s_2 + C_1 s_2 - C_2 s_2^3\right) = 0, \qquad (S40)$$

where $C_1$ and $C_2$ have been defined in Eq. (48), and by inserting Eqs. (S71) – (S72) into Eq. (48), these two coefficients have the following explicit forms:

$$C_1 = \mu\left(\lambda_1^2 + 2\lambda_2^2\right) + \mu\eta(i\omega)^{\beta_0}\left(\lambda_1^2 + \lambda_2^2 + \lambda_3^2\right), \qquad (S41)$$

$$C_2 = \mu\lambda_2^2 + \frac{1}{3}\mu\eta(i\omega)^{\beta_0}\left(\lambda_1^2 + \lambda_2^2 + \lambda_3^2\right). \qquad (S42)$$

$s_1$ and $s_2$ are the two roots solved by Eq. (41), and by inserting Eqs. (S71) – (S72) into Eq. (41), the quartic equation has the following explicit form:

$$\begin{aligned}C_2 s^4 + \left[\rho\frac{\omega^2}{k^2} - \mu\left(\lambda_1^2 + \lambda_2^2\right) - \frac{2}{3}\mu\eta(i\omega)^{\beta_0}\left(\lambda_1^2 + \lambda_2^2 + \lambda_3^2\right)\right]s^2 \\ + \mu\lambda_1^2 + \frac{1}{3}\mu\eta(i\omega)^{\beta_0}\left(\lambda_1^2 + \lambda_2^2 + \lambda_3^2\right) - \rho\frac{\omega^2}{k^2} = 0\end{aligned}. \qquad (S43)$$

## Supplementary Note 6. Derivation of secular equations of Lamb waves

### S6.1 Lamb waves in a fluid-immersed plate

The stream function of the plate is

$$\psi = \left[ A_1 \cosh(s_1 k x_2) + A_2 \sinh(s_1 k x_2) \right. \\ \left. + A_3 \cosh(s_2 k x_2) + A_4 \sinh(s_2 k x_2) \right] \exp\left[ i(k x_1 - \omega t) \right]. \quad (S44)$$

For the antisymmetric mode ($A_2 = A_4 = 0$), the stream function can be simplified as

$$\psi = \left[ A_1 \cosh(s_1 k x_2) + A_3 \cosh(s_2 k x_2) \right] \exp\left[ i(k x_1 - \omega t) \right]. \quad (S45)$$

For the symmetric mode ($A_1 = A_3 = 0$), the stream function can be simplified as

$$\psi = \left[ A_2 \sinh(s_1 k x_2) + A_4 \sinh(s_2 k x_2) \right] \exp\left[ i(k x_1 - \omega t) \right]. \quad (S46)$$

The potential function of the top fluid ($x_2 > h$) is

$$\varphi^+ = B_1 \exp(-\xi k x_2) \exp\left[ i(k x_1 - \omega t) \right], \quad (S47)$$

and the potential function of the bottom fluid ($x_2 < -h$) is

$$\varphi^- = B_2 \exp(\xi k x_2) \exp\left[ i(k x_1 - \omega t) \right]. \quad (S48)$$

The upper and lower surfaces of the plate are in contact with fluids, and satisfy the following fluid-solid interface conditions:

$$u_2 = u_2^f, \quad \Sigma_{21} = -\sigma_{22} u_{2,1}, \quad \Sigma_{22,1} = -p^f_{,1} - \sigma_{22} u_{2,12}, \quad \text{at} \quad x_2 = \pm h. \quad (S49)$$

For the antisymmetric mode, we can make use of the symmetry, therefore, only the boundary conditions at one side of the plate (e.g. $x_2 = h$) need to be considered. using $\psi$ defined by Eq. (S45) into boundary conditions, we obtain a system of linear equations

$$\mathbf{L}_{3\times 3}^{(\text{Lamb-F,A})} \left[ A_1, A_3, B_1 \right]^{\mathrm{T}} = 0, \quad (S50)$$

where the components of the matrix $\mathbf{L}_{3\times 3}^{(\text{Lamb-F,A})}$ include

$L_{11} = \cosh(s_1 k h), \quad L_{12} = \cosh(s_2 k h), \quad L_{13} = i\xi \exp(-\xi k h),$

$$L_{21} = (1+s_1^2)\cosh(s_1 kh), \quad L_{22} = (1+s_2^2)\cosh(s_2 kh), \quad L_{23} = 0,$$

$$L_{31} = \left(-\rho\frac{\omega^2}{k^2}s_1 + C_1 s_1 - C_2 s_1^3\right)\sinh(s_1 kh), \quad L_{32} = \left(-\rho\frac{\omega^2}{k^2}s_2 + C_1 s_2 - C_2 s_2^3\right)\sinh(s_2 kh),$$

$$L_{33} = i\rho^f \frac{\omega^2}{k^2}\exp(-\xi kh). \tag{S51}$$

where $C_1$ and $C_2$ have been defined in Eq. (48). $s_1$ and $s_2$ are the two roots solved by Eq. (41). $\xi$ is given in Supplementary Note 5.1 (as well as in Section 3.2). $\rho$ and $\rho^f$ denote the material densities of solid layer and fluid, respectively. $i$ in the elements $L_{13}$ and $L_{33}$ denotes the imaginary unit. To ensure the existence of non-trivial solutions in Eq. (S50), we have

$$\det\left(\mathbf{L}_{3\times 3}^{(\text{Lamb-F,A})}\right) = 0. \tag{S52}$$

By expanding Eq. (S52), the secular equation for the antisymmetric mode of fluid-immersed Lamb waves can be obtained (i.e. Eq. (51) in the main text).

For the symmetric mode, using $\psi$ defined by Eq. (S46) into boundary conditions, we obtain a system of linear equations

$$\mathbf{L}_{3\times 3}^{(\text{Lamb-F,S})}[A_2, A_4, B_1]^{\text{T}} = 0, \tag{S53}$$

where the components of the matrix $\mathbf{L}_{3\times 3}^{(\text{Lamb-F,S})}$ include

$$L_{11} = \sinh(s_1 kh), \quad L_{12} = \sinh(s_2 kh), \quad L_{13} = i\xi\exp(-\xi kh),$$

$$L_{21} = (1+s_1^2)\sinh(s_1 kh), \quad L_{22} = (1+s_2^2)\sinh(s_2 kh), \quad L_{23} = 0,$$

$$L_{31} = \left(-\rho\frac{\omega^2}{k^2}s_1 + C_1 s_1 - C_2 s_1^3\right)\cosh(s_1 kh), \quad L_{32} = \left(-\rho\frac{\omega^2}{k^2}s_2 + C_1 s_2 - C_2 s_2^3\right)\cosh(s_2 kh),$$

$$L_{33} = i\rho^f \frac{\omega^2}{k^2}\exp(-\xi kh). \tag{S54}$$

To ensure the existence of non-trivial solutions, we have

$$\det\left(\mathbf{L}_{3\times 3}^{(\text{Lamb-F,S})}\right) = 0. \tag{S55}$$

By expanding Eq. (S55), the secular equation for the symmetric mode of fluid-immersed Lamb waves can be obtained (i.e. Eq. (52) in the main text).

### S6.2 Lamb waves in a plate in vacuum

The boundary conditions of a plate in vacuum include:
$$\Sigma_{21} = 0, \quad \Sigma_{22,1} = 0, \text{ at } x_2 = \pm h. \tag{S56}$$

Applying Eqs. (S45), (S30) and (S31) into boundary conditions, we obtain a system of linear equations for the antisymmetric modes
$$\mathbf{L}_{2\times 2}^{(\text{Lamb,A})} [A_1, A_3]^{\text{T}} = 0, \tag{S57}$$

where the components of the matrix $\mathbf{L}_{2\times 2}^{(\text{Lamb,A})}$ include

$$L_{11} = (1+s_1^2)\cosh(s_1 kh), \quad L_{12} = (1+s_2^2)\cosh(s_2 kh),$$

$$L_{21} = \left(-\rho\frac{\omega^2}{k^2}s_1 + C_1 s_1 - C_2 s_1^3\right)\sinh(s_1 kh), \quad L_{22} = \left(-\rho\frac{\omega^2}{k^2}s_2 + C_1 s_2 - C_2 s_2^3\right)\sinh(s_2 kh). \tag{S58}$$

To ensure the existence of non-trivial solutions in Eq. (S57), we have
$$\det\left(\mathbf{L}_{2\times 2}^{(\text{Lamb,A})}\right) = 0. \tag{S59}$$

By expanding Eq. (S59), the secular equation for the antisymmetric mode of Lamb waves can be obtained (i.e. Eq. (53) in the main text).

Applying Eq. (S46), (S30) and (S31) into boundary conditions, we obtain a system of linear equations for the symmetric modes
$$\mathbf{L}_{2\times 2}^{(\text{Lamb,S})} [A_2, A_4]^{\text{T}} = 0, \tag{S60}$$

where the components of the matrix $\mathbf{L}_{2\times 2}^{(\text{Lamb,S})}$ include

$$L_{11} = (1+s_1^2)\sinh(s_1 kh), \quad L_{12} = (1+s_2^2)\sinh(s_2 kh),$$

$$L_{21} = \left(-\rho\frac{\omega^2}{k^2}s_1 + C_1 s_1 - C_2 s_1^3\right)\cosh(s_1 kh), \quad L_{22} = \left(-\rho\frac{\omega^2}{k^2}s_2 + C_1 s_2 - C_2 s_2^3\right)\cosh(s_2 kh). \tag{S61}$$

To ensure the existence of non-trivial solutions in Eq. (S60), we have

$$\det\left(\mathbf{L}_{2\times 2}^{(\text{Lamb},S)}\right) = 0. \tag{S62}$$

By expanding Eq. (S62), the secular equation for the symmetric mode of Lamb waves can be obtained (i.e. Eq. (54) in the main text).

### S6.3 Specific form of Lamb waves under the neo-Hookean model

Here we apply a specific constitutive model—neo-Hookean model, and present the corresponding secular equation of the Lamb waves of a plate in vacuum. The KVFD model is used to describe material viscoelasticity. Inserting explicit forms given by Eqs. (S71) – (S72) into Eq. (53), the secular equation of the antisymmetric modes can be simplified as

$$\begin{aligned} &\left(1+s_2^{\,2}\right)\cdot\left(-\rho\frac{\omega^2}{k^2}s_1 + C_1 s_1 - C_2 s_1^{\,3}\right)\cdot\tanh(s_1 kh) \\ &-\left(1+s_1^{\,2}\right)\cdot\left(-\rho\frac{\omega^2}{k^2}s_2 + C_1 s_2 - C_2 s_2^{\,3}\right)\cdot\tanh(s_2 kh) = 0 \end{aligned}, \tag{S63}$$

where $C_1$ and $C_2$ are defined by Eqs. (S41) and (S42), respectively. $s_1$ and $s_2$ are the two roots solved by Eq. (S43).

Similarly, the secular equation of the symmetric modes can be simplified from Eq. (54), which is written as

$$\begin{aligned} &\left(1+s_2^{\,2}\right)\cdot\left(-\rho\frac{\omega^2}{k^2}s_1 + C_1 s_1 - C_2 s_1^{\,3}\right)\cdot\coth(s_1 kh) \\ &-\left(1+s_1^{\,2}\right)\cdot\left(-\rho\frac{\omega^2}{k^2}s_2 + C_1 s_2 - C_2 s_2^{\,3}\right)\cdot\coth(s_2 kh) = 0 \end{aligned}. \tag{S64}$$

## S6.4 Specific form of fluid-immersed Lamb waves under the GOH model

In the following, we consider a specific constitutive model—GOH model, and present the corresponding secular equation of the Lamb waves of a fluid-immersed plate. The KVFD model is used to describe material viscoelasticity. Using the relation of Eq. (S82) (see Supplementary Note 7.4) into Eq. (51), the secular equation of the antisymmetric modes can be simplified as

$$\begin{aligned}&\left(1+s_2^2\right)\cdot\left(-\rho\frac{\omega^2}{k^2}s_1+C_1s_1-C_2s_1^3\right)\cdot\tanh(s_1kh)\\&-\left(1+s_1^2\right)\cdot\left(-\rho\frac{\omega^2}{k^2}s_2+C_1s_2-C_2s_2^3\right)\cdot\tanh(s_2kh)+\left(s_1^2-s_2^2\right)\frac{\rho^f}{\xi}\frac{\omega^2}{k^2}=0\end{aligned} \quad (S65)$$

where $C_1$ and $C_2$ are defined by

$$C_1 = 2G\beta + \gamma + \Omega \mathcal{A}_{03232} = 2\beta + \gamma + \eta(i\omega)^{\beta_0}(2\beta + \mathcal{A}_{03232}), \quad (S66)$$

$$C_2 = \gamma + \Omega Q = \gamma + \eta(i\omega)^{\beta_0} Q. \quad (S67)$$

$s_1$ and $s_2$ are the two roots solved by

$$C_2 s^4 + \left[\rho\frac{\omega^2}{k^2} - 2G\beta + \Omega(\alpha + \gamma - 2Q)\right]s^2 + \alpha + \Omega Q - \rho\frac{\omega^2}{k^2} = 0, \quad (S68)$$

where the explicit forms of $\alpha$, $\gamma$, $\beta$, $\mathcal{A}_{03232}$ are given by Eqs. (S80) when waves propagate along the axial direction, or Eq. (S81) when waves propagate along the circumferential direction. $Q$ is given by Eq. (S82). $G = 1 + \eta(i\omega)^{\beta_0}$ and $\Omega = \eta(i\omega)^{\beta_0}$.

Similarly, the secular equation of the symmetric modes can be simplified from Eq. (52), which gives

$$\begin{aligned}&\left(1+s_2^2\right)\cdot\left(-\rho\frac{\omega^2}{k^2}s_1+C_1s_1-C_2s_1^3\right)\cdot\coth(s_1kh)\\&-\left(1+s_1^2\right)\cdot\left(-\rho\frac{\omega^2}{k^2}s_2+C_1s_2-C_2s_2^3\right)\cdot\coth(s_2kh)+\left(s_1^2-s_2^2\right)\frac{\rho^f}{\xi}\frac{\omega^2}{k^2}=0\end{aligned}. \quad (S69)$$

**Supplementary Note 7. Explicit forms of incremental parameters for commonly used constitutive models**

Here, we present the explicit forms of incremental parameters for several commonly used constitutive models, including neo-Hookean model, Mooney-Rivlin model, Demiray-Fung model (Demiray, 1972), and GOH model (Gasser et al., 2006). These expressions facilitate the derivation of explicit relations for wave dispersion of plane shear waves, surface waves, and Lamb waves. It also serves as a reference for readers to conveniently select appropriate constitutive models and derive corresponding wave dispersion.

**S7.1 Neo-Hookean model**

The strain energy function is

$$W = \frac{\mu}{2}(I_1 - 3), \tag{S70}$$

where $\mu$ is the initial shear modulus of the material. Invariant $I_1 = \lambda_1^2 + \lambda_2^2 + \lambda_3^2$, where $\lambda_1$, $\lambda_2$, and $\lambda_3$ denote the stretch ratio along the $x_1$, $x_2$, and $x_3$ directions, respectively. The deformation gradient tensor $\boldsymbol{F} = \mathrm{diag}(\lambda_1, \lambda_2, \lambda_3)$. The general explicit forms of $\mathcal{A}_{0jikl}$ can be found in literature (Destrade, 2015). Applying the neo-Hookean model, the incremental parameters have the following explicit forms:

$$\alpha = \mu \lambda_1^2, \tag{S71-a}$$

$$\gamma = \mu \lambda_2^2, \tag{S71-b}$$

$$\beta = \frac{\mu}{2}(\lambda_1^2 + \lambda_2^2), \tag{S71-c}$$

$$\mathcal{A}_{01221} = \mathcal{A}_{02332} = 0, \tag{S71-d, e}$$

$$\mathcal{A}_{03232} = \mu \lambda_3^2. \tag{S71-f}$$

The explicit forms of stresses can be derived from Eqs. (33) - (35); they are:

$$\sigma^e_{D11} = \frac{1}{3}\mu\left(2\lambda_1^2 - \lambda_2^2 - \lambda_3^2\right), \tag{S72-a}$$

$$\sigma^e_{D22} = \frac{1}{3}\mu\left(2\lambda_2^2 - \lambda_1^2 - \lambda_3^2\right), \tag{S72-b}$$

$$Q = \frac{1}{3}\mu\left(\lambda_1^2 + \lambda_2^2 + \lambda_3^2\right). \tag{S72-c}$$

### S7.2 Mooney-Rivlin model

The strain energy function is

$$W = C_{10}(I_1 - 3) + C_{01}(I_2 - 3), \tag{S73}$$

where $C_{10}$ and $C_{01}$ are two constitutive parameters with the same dimensional units as stress. Invariants $I_1 = \lambda_1^2 + \lambda_2^2 + \lambda_3^2$, $I_2 = \lambda_1^2\lambda_2^2 + \lambda_1^2\lambda_3^2 + \lambda_2^2\lambda_3^2$. The incremental parameters have the following explicit forms:

$$\alpha = 2C_{10}\lambda_1^2 + 2C_{01}\lambda_2^{-2}, \tag{S74-a}$$

$$\gamma = 2C_{10}\lambda_2^2 + 2C_{01}\lambda_1^{-2}, \tag{S74-b}$$

$$\beta = C_{10}\left(\lambda_1^2 + \lambda_2^2\right) + C_{01}\left(\lambda_1^{-2} + \lambda_2^{-2}\right), \tag{S74-c}$$

$$\mathcal{A}_{01221} = -2C_{01}\lambda_3^{-2}, \tag{S74-d}$$

$$\mathcal{A}_{02332} = -2C_{01}\lambda_1^{-2}, \tag{S74-e}$$

$$\mathcal{A}_{03232} = 2C_{10}\lambda_3^2 + 2C_{01}\lambda_2^{-2}, \tag{S74-f}$$

The stresses have the following explicit forms:

$$\sigma^e_{D11} = \frac{2}{3}C_{10}\left(2\lambda_1^2 - \lambda_2^2 - \lambda_3^2\right) + \frac{2}{3}C_{01}\left(\lambda_2^{-2} + \lambda_3^{-2} - 2\lambda_1^{-2}\right), \tag{S75-a}$$

$$\sigma^e_{D22} = \frac{2}{3}C_{10}\left(2\lambda_2^2 - \lambda_1^2 - \lambda_3^2\right) + \frac{2}{3}C_{01}\left(\lambda_1^{-2} + \lambda_3^{-2} - 2\lambda_2^{-2}\right), \tag{S75-b}$$

$$Q = \frac{2}{3}C_{10}\left(\lambda_1^2 + \lambda_2^2 + \lambda_3^2\right) + \frac{4}{3}C_{01}\left(\lambda_1^{-2} + \lambda_2^{-2} + \lambda_3^{-2}\right). \tag{S75-c}$$

## S7.3 Demiray-Fung model

The Demiray–Fung model is widely used to describe biological soft tissues, as its exponential form can effectively capture the strain-stiffening behavior of fibers in tissues. The strain energy function is (Demiray, 1972)

$$W = \frac{\mu}{2b}\left\{\exp\left[b(I_1-3)\right]-1\right\}, \tag{S76}$$

where $\mu$ denotes the initial shear modulus of the material. $b$ (dimensionless) denotes the nonlinear stiffening effect of material. Invariant $I_1 = \lambda_1{}^2 + \lambda_2{}^2 + \lambda_3{}^2$. The incremental parameters have the following explicit forms:

$$\alpha = \mu e^{b(I_1-3)}\lambda_1{}^2, \tag{S77-a}$$

$$\gamma = \mu e^{b(I_1-3)}\lambda_2{}^2, \tag{S77-b}$$

$$\beta = \frac{\mu}{2}e^{b(I_1-3)}\left(\lambda_1{}^2+\lambda_2{}^2\right) + \mu b e^{b(I_1-3)}\left(\lambda_1{}^2-\lambda_2{}^2\right)^2, \tag{S77-c}$$

$$\mathcal{A}_{01221} = \mathcal{A}_{02332} = 0, \tag{S77-d, e}$$

$$\mathcal{A}_{03232} = \mu e^{b(I_1-3)}\lambda_3{}^2, \tag{S77-f}$$

The stresses have the following explicit forms:

$$\sigma^e_{D11} = \frac{1}{3}\mu e^{b(I_1-3)}\left(2\lambda_1{}^2 - \lambda_2{}^2 - \lambda_3{}^3\right), \tag{S78-a}$$

$$\sigma^e_{D22} = \frac{1}{3}\mu e^{b(I_1-3)}\left(2\lambda_2{}^2 - \lambda_1{}^2 - \lambda_3{}^3\right), \tag{S78-b}$$

$$Q = \frac{1}{3}\mu e^{b(I_1-3)}\left(\lambda_1{}^2 + \lambda_2{}^2 + \lambda_3{}^2\right). \tag{S78-c}$$

## S7.4 Gasser-Ogden-Holzapfel model

The Gasser-Ogden-Holzapfel (GOH) model has been widely used to describe arteries (Gasser et al., 2006). The strain energy function is (i.e. Eq. (55) in the main text):

$$W = \frac{\mu}{2}(I_1 - 3) + \frac{k_1}{2k_2} \sum_{i=4,6} \left\{ \exp\left[ k_2 \left( \kappa I_1 + (1 - 3\kappa) I_i - 1 \right)^2 \right] - 1 \right\}, \quad \text{(S79)}$$

where $\mu$ and $k_1$ denote the initial shear modulus of elastin and collagen fibers, respectively. $k_2$ (dimensionless) denotes the nonlinear stiffening of collagen fibers. $\kappa$ represents the fiber dispersion ($0 \leq \kappa \leq 1/3$). Invariants $I_1 = \text{tr}(\boldsymbol{C})$, $I_4 = \boldsymbol{M} \cdot \boldsymbol{CM}$ and $I_6 = \boldsymbol{M'} \cdot \boldsymbol{CM'}$. $\boldsymbol{C}$ is the right Cauchy-Green strain tensor. $\boldsymbol{M}$ and $\boldsymbol{M'}$ denote two symmetrically distributed fiber orientations. $\phi$ denotes the angle between the fiber orientation and the circumferential direction (Fig. S2).

In the first case, where the $x_1$ axis is aligned with the axial direction of the artery sample (Fig. S2a), the incremental parameters have the following explicit forms:

$$\alpha = 2W_1 \lambda_1^2 + 2W_4 \lambda_1^2 \sin^2 \phi + 2W_6 \lambda_1^2 \sin^2 \phi, \quad \text{(S80-a)}$$

$$\gamma = 2W_1 \lambda_2^2, \quad \text{(S80-b)}$$

$$\begin{aligned}
\beta &= W_1 \left( \lambda_1^2 + \lambda_2^2 \right) + W_4 \lambda_1^2 \sin^2 \phi + W_6 \lambda_1^2 \sin^2 \phi + 2W_{11} \left( \lambda_1^2 - \lambda_2^2 \right)^2 \\
&+ 4W_{14} \lambda_1^2 \left( \lambda_1^2 - \lambda_2^2 \right) \sin^2 \phi + 4W_{16} \lambda_1^2 \left( \lambda_1^2 - \lambda_2^2 \right) \sin^2 \phi \\
&+ 2W_{44} \lambda_1^4 \sin^4 \phi + 2W_{66} \lambda_1^4 \sin^4 \phi
\end{aligned} \quad \text{(S80-c)}$$

$$\mathcal{A}_{01221} = \mathcal{A}_{02332} = 0, \quad \text{(S80-d, e)}$$

$$\mathcal{A}_{03232} = 2W_1 \lambda_3^2 + 2W_4 \lambda_3^2 \cos^2 \phi + 2W_6 \lambda_3^2 \cos^2 \phi, \quad \text{(S80-f)}$$

where $W_i = \partial W / \partial I_i$, $W_{ij} = \partial^2 W / \partial I_i \partial I_j$. $I_1 = \lambda_1^2 + \lambda_2^2 + \lambda_3^2$, $I_4 = I_6 = \lambda_1^2 \sin^2 \phi + \lambda_3^2 \cos^2 \phi$. When $\lambda_1 = \lambda_2 = \lambda_3 = 1$, Eq. (S80) reduces to $\alpha = \gamma = \mathcal{A}_{03232} = \mu$, and $\beta = \mu + 4k_1 (1 - 3\kappa)^2 \sin^4 \phi$.

In the second case, where the $x_1$ axis is aligned with the circumferential direction of the artery sample (Fig. S2b), the incremental parameters have the following explicit forms:

$$\alpha = 2W_1 \lambda_1^2 + 2W_4 \lambda_1^2 \cos^2 \phi + 2W_6 \lambda_1^2 \cos^2 \phi, \quad \text{(S81-a)}$$

$$\gamma = 2W_1 \lambda_2^2, \quad \text{(S81-b)}$$

$$\begin{aligned}
\beta &= W_1 \left( \lambda_1^2 + \lambda_2^2 \right) + W_4 \lambda_1^2 \cos^2 \phi + W_6 \lambda_1^2 \cos^2 \phi + 2W_{11} \left( \lambda_1^2 - \lambda_2^2 \right)^2 \\
&+ 4W_{14} \lambda_1^2 \left( \lambda_1^2 - \lambda_2^2 \right) \cos^2 \phi + 4W_{16} \lambda_1^2 \left( \lambda_1^2 - \lambda_2^2 \right) \cos^2 \phi \\
&+ 2W_{44} \lambda_1^4 \cos^4 \phi + 2W_{66} \lambda_1^4 \cos^4 \phi
\end{aligned} \quad \text{(S81-c)}$$

$$\mathcal{A}_{01221} = \mathcal{A}_{02332} = 0, \quad \text{(S81-d, e)}$$

$$\mathcal{A}_{03232} = 2W_1 \lambda_3^2 + 2W_4 \lambda_3^2 \sin^2 \phi + 2W_6 \lambda_3^2 \sin^2 \phi. \quad \text{(S81-f)}$$

where $W_i = \partial W / \partial I_i$, $W_{ij} = \partial^2 W / \partial I_i \partial I_j$. $I_1 = \lambda_1^2 + \lambda_2^2 + \lambda_3^2$, $I_4 = I_6 = \lambda_1^2 \cos^2\phi + \lambda_3^2 \sin^2\phi$. When $\lambda_1 = \lambda_2 = \lambda_3 = 1$, Eq. (S81) reduces to $\alpha = \gamma = \mathcal{A}_{03232} = \mu$, and $\beta = \mu + 4k_1(1-3\kappa)^2 \cos^4\phi$.

In both cases, the stresses can be calculated in a consistent form as follows:

$$\sigma^e_{D11} = \frac{2}{3}\alpha - \frac{1}{3}\gamma - \frac{1}{3}\mathcal{A}_{03232}, \tag{S82-a}$$

$$\sigma^e_{D22} = \frac{2}{3}\gamma - \frac{1}{3}\alpha - \frac{1}{3}\mathcal{A}_{03232}, \tag{S82-b}$$

$$Q = \frac{1}{3}(\alpha + \gamma + \mathcal{A}_{03232}). \tag{S82-c}$$

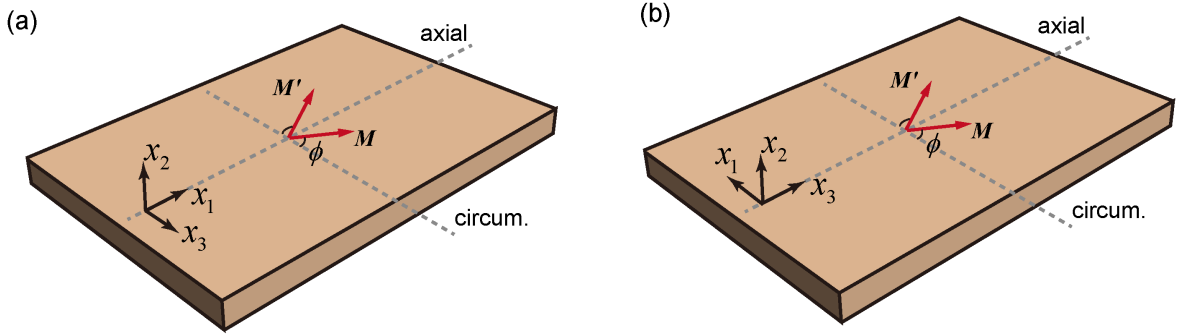

**Figure S2.** Schematic of the Gasser-Ogden-Holzapfel model to describe arteries. **(a)** The $x_1$ axis is aligned with the axial direction of the artery sample. **(b)** The $x_1$ axis is aligned with the circumferential direction of the artery sample.

**Supplementary Note 8.** *Ex vivo* **experiments of porcine ascending aortas**

**S8.1 Tensile test to measure hyperelastic parameters of the sample**

In order to characterize hyperelastic (constitutive) parameters of ascending aortas, the aorta sample was tested by a uniaxial tensile (ElectroForce 3200, TA Instruments, USA). The GOH model was employed to describe the artery sample, and the strain energy function $W$ have been presented in Eq. (55) and Eq. (S79). The sample is stretched along the $x_1$ direction (circumferential), with $x_2$ denoting the wall thickness direction, and $x_3$ denoting the axial direction of the sample. In the case of uniaxial stretch, the relationship of stresses and stretch ratios satisfies (Ogden, 2003)

$$\sigma_1 = \lambda_1 \frac{\partial W}{\partial \lambda_1} - \lambda_3 \frac{\partial W}{\partial \lambda_3}, \tag{S83}$$

$$\lambda_2 \frac{\partial W}{\partial \lambda_2} = \lambda_3 \frac{\partial W}{\partial \lambda_3}, \tag{S84}$$

together with the incompressible condition $\lambda_1 \lambda_2 \lambda_3 = 1$.

Combining Eqs. (S83) and (S84), we can calculate the theoretical relationship between $\sigma_1$ and $\lambda_1$. To fit constitutive parameters, we define the loss function $\mathcal{F}$ as

$$\mathcal{F} = \sqrt{\frac{\sum_{j=1}^{n}\left(\sigma^{(j,\text{theo})} - \sigma^{(j,\text{exp})}\right)^2}{n}}, \tag{S85}$$

where $\sigma^{(j,\text{theo})}$ ($j$ = 1, 2, …, $n$, $n$ is the total number of data points) denotes the theoretically predicted stress ($\sigma_1$). $\sigma^{(j,\text{exp})}$ denotes the experimentally measured stress. The genetic algorithm was employed to minimize the loss function and search the best-fit values. Figure S3 shows the experimental stretch-stress curve and the fitting curve. The fitting parameters are $\mu = 33.4 \text{ kPa}$, $k_1 = 72.7 \text{ kPa}$, $k_2 = 6.3$, $\kappa = 0.26$, $\phi = 42.8°$.

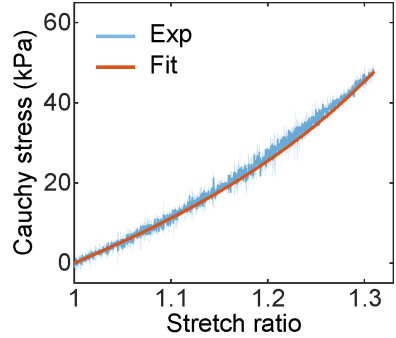

**Figure S3**. Stretch-stress data obtained by the uniaxial tensile test along the circumferential direction of the artery sample.

### S8.2 Parameter sensitivity analysis of the stretch-strain curve

The parameter sensitivity analysis was performed to study the uncertainty of the fitting parameters. Each constitutive parameter was varied by ±20% around its fitted value, and the corresponding stretch-stress curves were theoretically calculated and plotted in Fig. S4. The relative changes of the curves related to parameters $\mu$, $\kappa$ and $\phi$ are significant, indicating that these parameters can be fitted in a stable and reliable way, while the uncertainty of $k_2$ might be relatively obvious.

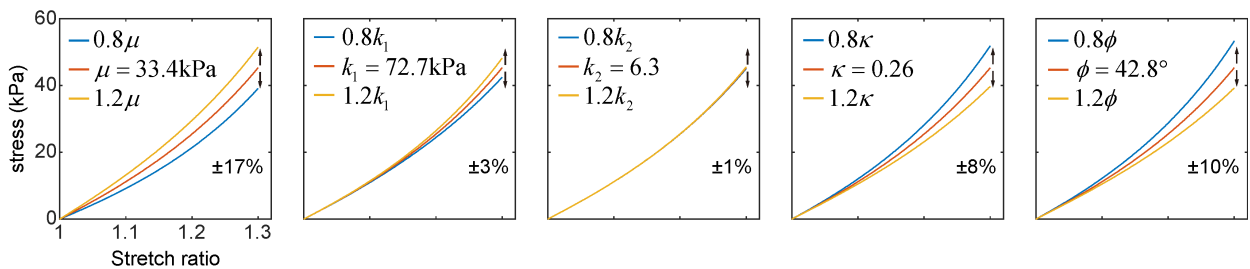

**Figure S4**. Parameter sensitivity analysis of GOH parameters on the stretch-stress curve, from left to right: $\mu$, $k_1$, $k_2$, $\kappa$, and $\phi$.

## S8.3 Parameter sensitivity analysis of the wave dispersion

We further discuss the parameter sensitivity of the KVFD parameters on the dispersion curve. As shown in Fig. S5, the relative change of the curve with varying $\beta_0$ is obvious (>10%), while the relative change related to $\eta$ is lower than 5%. This suggests that the fractional order $\beta_0$ can be inversed more reliably than the parameter $\eta$.

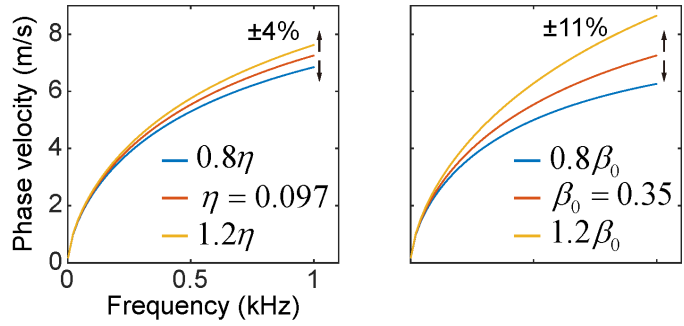

**Figure S5**. Parameter sensitivity analysis of the KVFD parameters $\eta$ and $\beta_0$ on the wave dispersion curve (A0 mode of Lamb waves).

**Supplementary Note 9. Extraction of the phase velocity and attenuation in the finite element analysis**

Here, we provide details about the extraction of phase velocity and wave attenuation from the finite element results. Fig. S6a shows the particle velocity field of the finite element model at a specific time, where the bulk material was firstly pre-stretched in $x_1$ and then a sinusoidal force was applied on the model along $x_2$ and $x_1$ respectively (see Section 5.3). By extracting the particle velocity along the path of plane wave propagation, we obtained a spatio-temporal velocity field as shown in Fig. S6b. The phase velocity $c$ was measured by fitting slope in the map, i.e., $c = \Delta L / \Delta t$. In order to get attenuation, we extracted the velocity amplitude along wave path at a specific time, and the attenuation was obtained by a linear fitting of the curve of the logarithm of the normalized velocity with respect to length, i.e., $k_{\text{im}} = \Delta \ln(v_0 / v_{\text{max}}) / \Delta L$ (Fig. S6c).

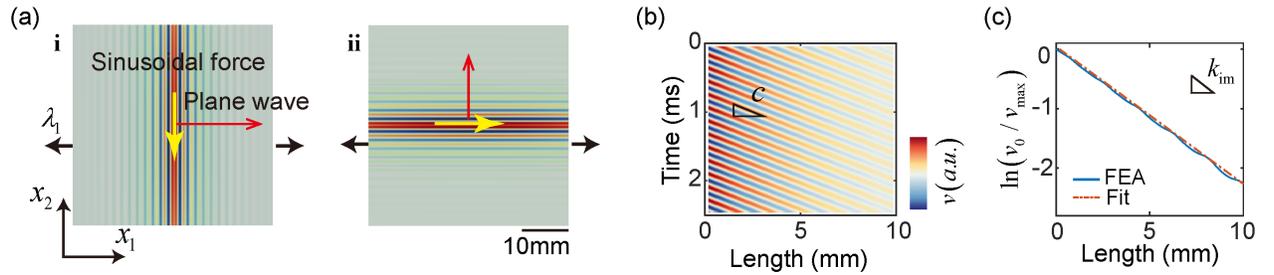

**Figure S6**. Extraction of phase velocity and attenuation from the simulation results. **(a)** finite element models. **i**. The sinusoidal force is applied along $x_2$ and the plane shear wave propagates along $x_1$. **ii**. The force is applied along $x_1$ and the wave propagates along $x_2$. **(b)** Spatio-temporal velocity field of plane waves. The phase velocity $c$ is fitted therein. **(c)** Log-scale of the normalized particle velocity versus length. The attenuation $k_{\text{im}}$ is obtained from the slope of the fitted curve.